\documentclass{IEEEtran}
\usepackage[tbtags]{amsmath}
\usepackage{cite}
\usepackage{amsmath,amssymb,amsfonts}
\usepackage{algorithmic}
\usepackage{graphicx}

\usepackage{textcomp}
\usepackage{multirow}
\usepackage{hyperref}
\usepackage{epstopdf}
\usepackage{subcaption}
\usepackage{tabularx}
\usepackage{color}
\usepackage{array}
\usepackage{booktabs}
\usepackage{stfloats}
\usepackage{url}
\usepackage{verbatim}
\usepackage{makecell}
\usepackage{multirow}
\usepackage{threeparttable}
\usepackage{subcaption} 
\usepackage{float}
\usepackage{fancyhdr}
\usepackage[table]{xcolor} 
\def\BibTeX{{\rm B\kern-.05em{\sc i\kern-.025em b}\kern-.08em
    T\kern-.1667em\lower.7ex\hbox{E}\kern-.125emX}}

\fancypagestyle{IEEEtitlepagestyle}{%
    \fancyhf{}
    \fancyhead[L]{\normalsize S. Sun et al., “Modeling and Analysis of Land-to-Ship Maritime Wireless Channels at 5.8 GHz," \textit{ IEEE Transactions on Wireless Communications}, doi: 10.1109/TWC.2025.3634176.}
    
}
\fancypagestyle{plain}{%
    \fancyhf{}

}

\begin{document}
\title{\textcolor{black}{Modeling and Analysis of Land-to-Ship Maritime Wireless Channels at 5.8 GHz}}
\author{Shu Sun, \IEEEmembership{Senior Member, IEEE}, Yulu Guo, Meixia Tao, \IEEEmembership{Fellow, IEEE}, Wei Feng, \IEEEmembership{Senior Member, IEEE}, \\
	Jun Chen, Ruifeng Gao, \IEEEmembership{Member, IEEE}, Ye Li, \IEEEmembership{Member, IEEE}, Jue Wang, \IEEEmembership{Member, IEEE}, and \\
	Theodore S. Rappaport, \IEEEmembership{Fellow, IEEE}

\thanks{S. Sun, Y. Guo, and M. Tao are with the School of Information Science and Electronic Engineering, Shanghai Jiao Tong University, Shanghai 200240, China (e-mail: \{shusun, guoyl1, mxtao\}@sjtu.edu.cn).

W. Feng and J. Chen are with the Department of Electronic Engineering, Tsinghua University, Beijing 100084, China (e-mail: \{fengwei, junch\}@tsinghua.edu.cn).

R. Gao is with the School of Transportation and Civil Engineering, Nantong University, and is also with Nantong Research Institute for Advanced Communication
Technologies, Nantong 226019, China (e-mail: grf@ntu.edu.cn).

Y. Li and J. Wang are with the School of Information Science and Technology, Nantong University, and are also with Nantong Research Institute for Advanced Communication
Technologies, Nantong 226019, China (e-mail: \{yeli, wangjue\}@ntu.edu.cn).

T. S. Rappaport is with the NYU WIRELESS research center and Tandon School of Engineering, New York University, Brooklyn, NY 11201, USA (e-mail: tsr@nyu.edu).

}
}

\maketitle

\begin{abstract}
 
 \textcolor{black}{Maritime channel modeling is crucial for designing robust nearshore communication systems, yet reliable models that account for the dynamic marine environment with varying sea waves, wind conditions, and vessel motions remain scarce.} This article investigates land-to-ship maritime wireless channel characteristics at 5.8 GHz based upon an extensive measurement campaign, with concurrent hydrological and meteorological information collection. First, a novel large-scale path loss model with physical foundation and high accuracy is proposed for dynamic marine environments. Then, we introduce the concept of sea-wave-induced fixed-point (SWIFT) fading, a peculiar phenomenon in maritime scenarios that captures the impact of sea surface fluctuations on received power. An enhanced two-ray model incorporating vessel rotational motion is propounded to simulate the SWIFT fading, showing good alignment with measured data, particularly for modest antenna movements. Next, the small-scale fading is studied by leveraging a variety of models including the two-wave with diffuse power (TWDP) and asymmetric Laplace distributions, with the latter performing well in most cases, while TWDP better captures bimodal fading in rough seas. Furthermore, maritime channel sparsity is examined via the Gini index and Rician $K$ factor, and temporal dispersion is characterized. The resulting channel models and parameter characteristics offer valuable insights for maritime wireless system design and deployment.
\end{abstract}


\section{Introduction}
\label{sec:introduction}
\IEEEPARstart{I}{n} recent decades, the rapid growth of the marine economy and the surge in offshore activities have heightened the demand for reliable, high-speed land-to-ship communication. Consequently, the development of maritime wireless communication networks has become a critical focus. Within this process, a thorough understanding of maritime wireless channel characteristics serves as a foundational and indispensable component. In the existing maritime communication networks, satellite-based systems can provide high data rate communication services, but they are unavoidably associated with high service costs and substantial transmission delays \cite{Alqurashi23ITJ}. Currently, efforts are focused on achieving wide-area coverage for nearshore users through terrestrial base stations or maritime relay platforms\footnote{The nearshore usually denotes the realm within 20 nautical miles (about 37 km) of the shoreline~\cite{COIN}.} \cite{Wang18Access}. These base stations and platforms primarily operate in the medium frequency, high frequency, and very high frequency (VHF) ranges~\cite{Wang18Access}, which imposes limitations on site selection and communication bandwidths\footnote{For instance, only about 6 MHz bandwidth is allocated for the VHF maritime mobile band as specified by the International Telecommunication Union (ITU)\cite{ITU2231}.}, making it desirable to utilize higher frequency bands with more abundant spectrum allocations. Thus, it is essential to conduct more measurements to investigate and model maritime channel characteristics in higher frequency bands, since channel models are indispensable cornerstones for network design, evaluation, and deployment. 


Channel modeling typically begins with measurements in various representative environments to extract key channel parameters and develop corresponding models, including standardized or widely adopted models such as WINNER \uppercase\expandafter{\romannumeral2}\cite{Kyosti07WINNER}, 3rd Generation Partnership Project (3GPP)\cite{3GPP}, the Simulation of Indoor Radio Channel Impulse Response (CIR) Models (SIRCIM)\cite{Rappaport91TC}, the Simulation of Mobile Radio CIR Models (SMRCIM)\cite{Rappaport93TC}, and NYUSIM\cite{Poddar24CST}. Numerous channel measurements and models exist for 5G/6G terrestrial scenarios, e.g., \cite{Jiang21Net},\cite{Wang20VTM},\cite{Cai24CM},\cite{Sun18TVT},\cite{Jiang20TC} and references therein, which offer insightful perspectives on terrestrial channel modeling and characteristics analysis. However, when considering the extension of terrestrial broadband mobile communication systems to nearshore areas, the unique attributes of the maritime environment necessitate new channel measurement and modeling efforts. Unlike terrestrial scenarios, the maritime radio propagation environment is significantly affected by various factors such as wind speed, temperature, sea states, and vessel motion. Additionally, scatterers and users are usually sparsely distributed over the vast sea surface, rendering traditional assumptions like Rayleigh fading invalid. Instead, maritime wireless channels exhibit sparsity and location dependency\cite{Wang18Access}. Furthermore, sea surface fluctuations are time-varying and not confined to fixed temporal scales, leading to more complex random undulations in nearshore wireless channels \cite{He22TCommun}. Thus, accurate modeling must account for the stochastic effects of wave dynamics.

Among the existing maritime channel measurements, researchers mainly focused on the S-band and C-band. Lee \emph{et al.} conducted land-to-ship channel measurements in the Yellow Sea of China, analyzing the received signal strength and the small-scale fading in the line-of-sight (LoS) propagation region at 2.4 GHz, and the results indicated that small-scale fading is best characterized by the Rician distribution \cite{Lee17WCL}. Authors in \cite{Yang11VTC} investigated the applicability of several terrestrial wireless channel models in maritime environments at 2 GHz and found that the ITU-R P.1546-2 model\cite{ITU-RP.1546} showed the best agreement with the measured results. By adding a Gaussian random variable to the length of the rays in the traditional two-ray model, the work \cite{Dahman19IWCL} described the randomness introduced by environmental changes in maritime channels. The researchers in \cite{Mi25TAP} conducted land-to-ship channel measurements at 3.2 GHz and analyzed how islands affect wireless propagation, revealing that cluster delays shifted linearly with distance and small-scale fading changed from Rician to log-normal distribution when the ship passed through islands. In \cite{Lee14Radioengineering}, measurements were conducted at 5 GHz in tropical sea regions, and it was proposed that a third refracted path, in addition to the direct and reflected paths, may occur during long-distance (beyond the break-point distance in the two-ray model) maritime transmission due to the evaporation duct effect over the sea. The measurement results showed excellent agreement with the theoretical curves \cite{Lee14Radioengineering}, indicating that considering the duct effect is necessary in long-distance maritime scenarios. In\cite{Reyes-Guerrero16Ocean}, channel measurements were carried out at 5.8 GHz and instantaneous power delay profiles (PDPs) of wireless links were acquired between buoys and ships under various multipath conditions, followed by the investigation on delay dispersion characteristics. In \cite{Joe07WCNC}, 5.8 GHz fixed WiMAX measurements were conducted in harbor environments to study the impacts of multipath effects, Doppler shift, and vessel motion on wireless channels. \textcolor{black}{For the same frequency band, a land-to-ship channel measurement campaign was launched~\cite{Wei21Tsinghua,Wei21IoTJ}, accompanied with some preliminary analysis on path loss, small-scale fading, and root-mean-square (RMS) delay spread.} Ship-to-ship measurements at 5.9 GHz with a maximum distance of 2.6 km were performed\cite{Yang18ITST}, where the statistical properties of several channel parameters were analyzed. The authors of~\cite{Zhang23TAP} conducted measurements with different receiver (Rx) heights at 8 GHz in the South China Sea. Among the fitting results of various path loss models, the log-distance model exhibited the smallest root mean square error (RMSE), while the parabolic equation model provided additional information on interference nulls. For small-scale fading, the Rician distribution showed the best Kolmogorov–Smirnov (K-S) test metrics compared to other commonly used distributions. 

Attention is also gradually shifting towards even higher frequency bands. A beyond-the-horizon maritime channel measurement campaign was conducted at 25 GHz for fixed-to-fixed point links\cite{Lyu24AWPL}. Based on the measurement data, the authors analyzed the short-term variations in received signal power caused by environmental changes and presented empirical models. To better align with real-world conditions, some studies have applied modifications to the traditional two-ray model to more comprehensively account for fading produced by maritime environmental factors. For instance, \cite{Zhao13ISCCC}
combined reflection from the sea surface and antenna height variation to propose a modified two-ray model through theoretical analysis, while \cite{Mehrnia16ICST} introduced an exponential coefficient into the two-ray model formula to correct for fluctuations affected by environmental factors via ray tracing, and achieved better prediction results in the millimeter-wave frequency band. 

To date, the existing works on maritime channel measurements and modeling have not considered the impact of various environmental factors on channel characteristics, or the investigated distances are limited, or the channel parameters and models are not sufficiently comprehensive. In this paper, real-world channel measurement data at 5.8 GHz for the land-to-ship scenario in the Yellow Sea of China~\cite{Wei21Tsinghua,Wei21IoTJ} is used to conduct systematic modeling and analysis on a variety of key channel parameters such as large-scale path loss, power fading on different time scales, channel sparsity, and temporal statistics. The main novelty and contributions of this work, \textcolor{black}{compared with the current literature including our previous studies~\cite{Wei21Tsinghua} and~\cite{Wei21IoTJ},} are as follows:
\begin{itemize}
    \item Environmental information is taken into account when carrying out the channel \textcolor{black}{modeling and analysis}, and variations in channel characteristics under different sea conditions are compared and discussed in detail. Results demonstrate that the environment has a significant impact on diverse channel parameters.
    \item A series of large-scale fading models are employed to characterize the location-dependent path loss. More importantly, a new \textcolor{black}{dual-slope} large-scale path loss model is propounded, which embodies a solid physical basis, high modeling accuracy, and super adaptability over a wide range of distances and sea conditions. In addition, the two-wave with diffuse power (TWDP) model that has Rayleigh and Rician distributions as special cases \cite{Durgin02ITC} and asymmetric Laplace function are utilized to depict the small-scale fading, in addition to the conventional Rician, Nakagami-m, and lognormal distributions. 
    \item We introduce the concept of sea-wave-induced fixed-point (SWIFT) fading to describe the received power fading caused by fluctuations of sea waves \textcolor{black}{and three-dimensional (3D) vessel swaying} while the vessel stays at a fixed coordinate, which is a distinctive feature in maritime channels as compared with terrestrial scenarios. An enhanced two-ray model based on a wave fluctuation spectrum is developed, \textcolor{black}{in conjunction with 3D vessel swinging models, to characterize SWIFT fading. The proposed model well matches measured data under modest antenna displacements.}
    \item Channel sparsity for the marine propagation environment is investigated in depth, where different metrics including the Gini index and Rician $K$ factor are adopted to quantify the sparsity level. The Gini index stems from economics \cite{Hurley09TIT} and has been shown to predict channel sparsity accurately, and it is garnering attention recently in research on land channels but has not been considered for maritime scenarios so far to our best knowledge.   
\end{itemize}

The remainder of this paper is organized as follows. In Section~\ref{sec_mea}, the maritime wireless channel measurement system and procedure are presented, along with a brief description of the data processing method. Sections~\ref{sec_PL} to~\ref{sec_small-scale} focus on the large-scale, SWIFT, and small-scale fading, respectively. Channel sparsity is analyzed in Section~\ref{sec_sparsity}, followed by the temporal statistics in Section~\ref{sec_temp}. Conclusions are drawn in Section~\ref{sec_con}.

\section{Measurement Campaign}\label{sec_mea}
The land-to-ship maritime channel measurements were conducted from the Qidong Campus of Nantong University to the Qidong waters of the Yellow Sea in 2018~\cite{Wei21Tsinghua,Wei21IoTJ}. In this section, we provide the necessary details of the measurements and data processing in order to facilitate the in-depth investigation of channel characteristics in the later sections. 

\subsection{Measurement System}
\begin{figure}[!t]
    \centerline{\includegraphics[width=0.85\columnwidth]{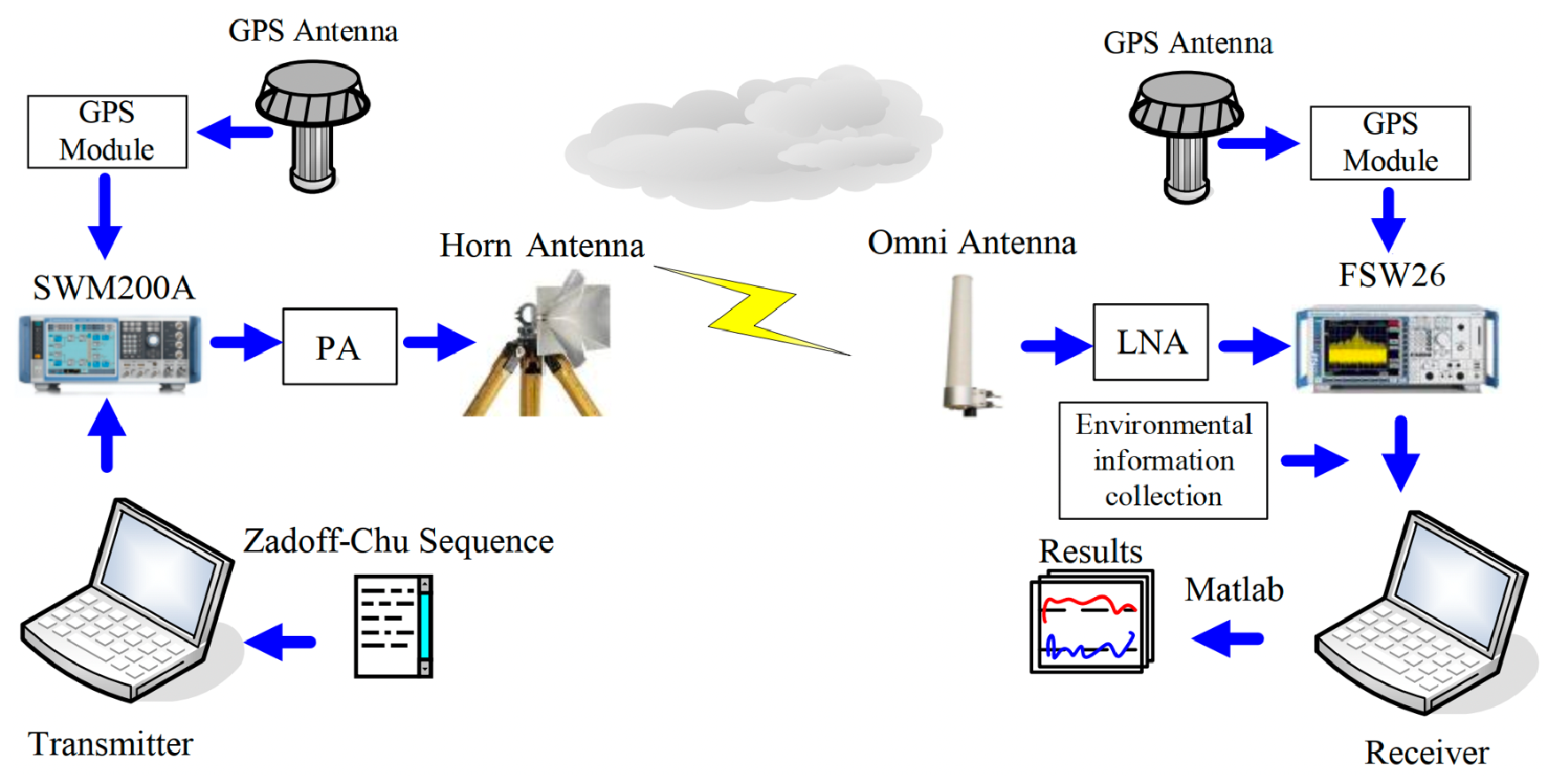}}
    \caption{\textcolor{black}{Diagram of channel measurement system (re-drawing of Fig. 1 in~\cite{Wei21Tsinghua}).}}
    \label{fig1}
\end{figure}

The measurement system is sketched in Fig. \ref{fig1}. The transmitter (Tx) utilizes a Zadoff-Chu (ZC) sequence, known for its excellent autocorrelation properties, as the channel sounding signal, with a length of 65535. The vector signal generator (R\&S® SMW200A) is used to perform binary phase shift keying modulation and digital-to-analog conversion. The generated signal, with a carrier frequency of 5.8 GHz and a bandwidth of 20 MHz, is amplified by a power amplifier with a gain of 30 dB and then transmitted via a horn antenna with a half-power beamwidth (HPBW) of 65° and a boresight gain of 9 dBi to provide directional gain~\cite{Rappaport12RWS}.

\begin{figure}[!t]
    \centerline{\includegraphics[width=0.8\columnwidth, height=8cm]{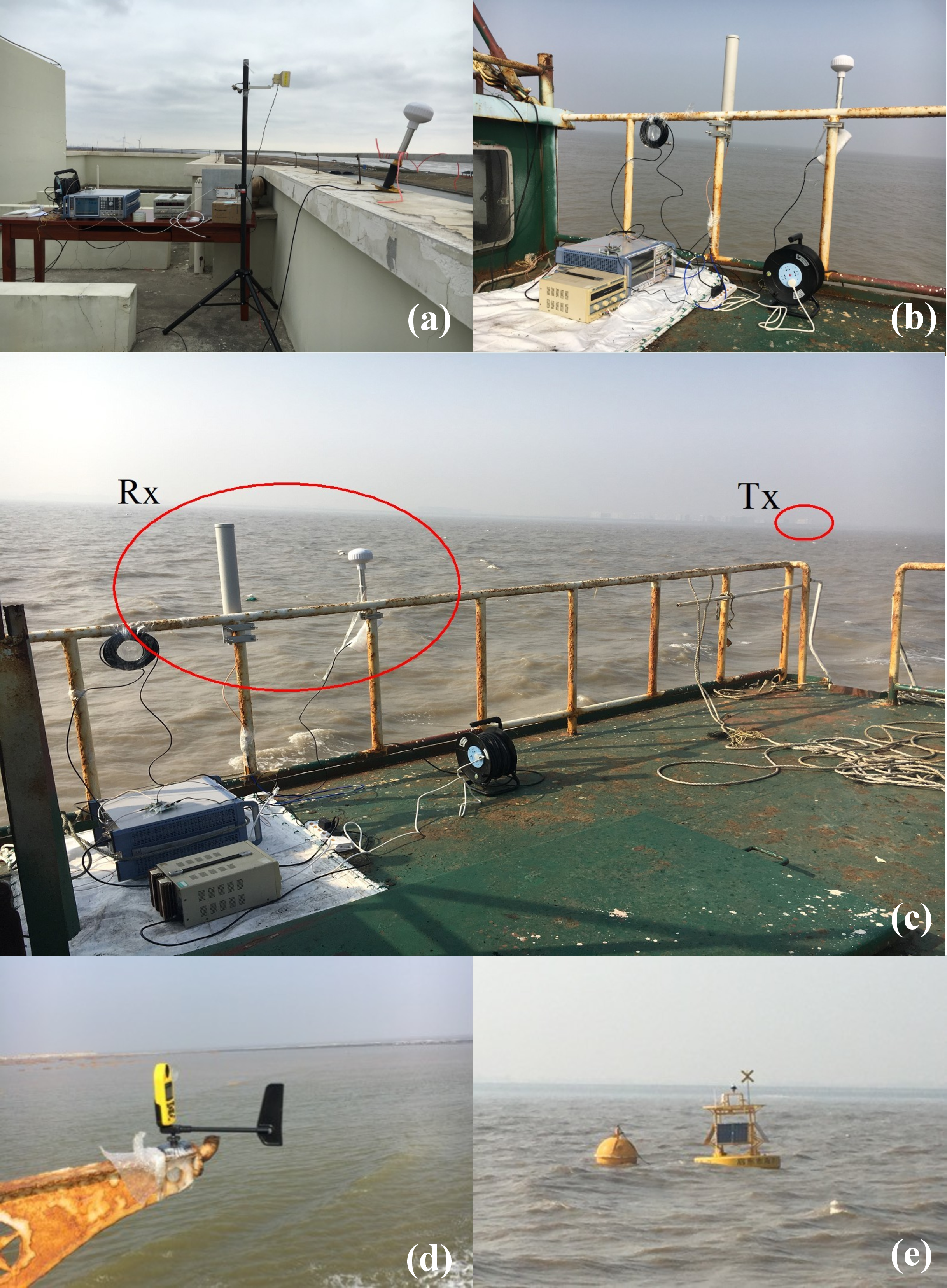}}
    \caption{Snapshots of measurement system and scenario. (a) Tx equipment. (b) Rx equipment. (c) Measurement scenario over the open sea. (d) Weather meter. (e) Monitoring buoy in the sea area of Qidong City.}
    \label{fig2}
\end{figure}

At the Rx, also shown in Fig. \ref{fig1}, the signal is received by an omnidirectional antenna with a gain of 13 dBi. The received signal is then amplified by a low noise amplifier with 27.4 dB gain and subsequently processed by a spectrum and signal analyzer (R\&S® FSW43) for analog-to-digital conversion and IQ demodulation. A 10 MHz GPS signal is employed at both the Tx and Rx to synchronize their respective transmission and reception times. \textcolor{black}{Environmental information is collected along with the measured data.} Fig. \ref{fig2} presents photographs of the measurement equipment and the overall measurement scenario over the open sea. More details of the measurement system parameter settings are listed in Table \ref{tab:system_parameters}.

\begin{figure}[!t]
    \centering
    \includegraphics[width=0.8\columnwidth]{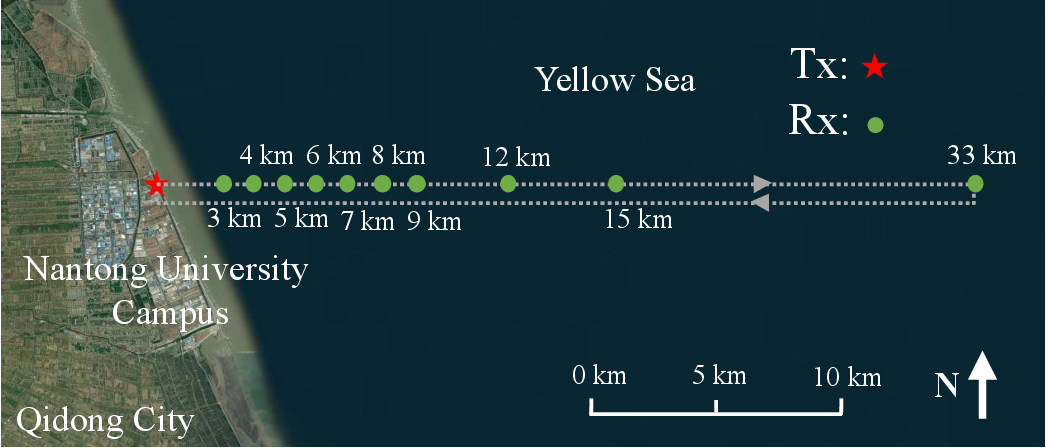}
    \caption{Tx location and Rx trajectory in the maritime channel measurements.}
    \label{fig3}
\end{figure}

\begin{table}[h!]
\caption{\textsc{Measurement System Parameters}}
\centering
\begin{tabular}{|c|c|}
\hline
\textbf{Parameter} & \textbf{Value} \\
\hline\hline
Carrier Frequency & 5.8 GHz \\
\hline
Bandwidth & 20 MHz \\
\hline
Tx Antenna Height & 25 m \\
\hline
\multirow{2}{*}{Rx Antenna Height} & High: 4 m \\
\cline{2-2}
& Low: 1 m \\
\hline
Transmit Power & Day 1: 15 dBm; Day 2: 25 dBm \\
\hline
\multirow{3}{*}{Tx Antenna Type} & Horn Antenna \\
\cline{2-2}
& HPBW: 65° (Azimuth and Elevation) \\
\cline{2-2}
& Antenna Gain: 9 dBi \\
\hline
\multirow{2}{*}{Rx Antenna Type} & Omnidirectional Antenna \\
\cline{2-2}
& Antenna Gain: 13 dBi \\
\hline
Tx Power Amplifier & 30 dB \\
\hline
Rx Low Noise Amplifier & 27.4 dB \\
\hline
Line Loss & $\sim$ 1 dB \\
\hline
Baseband Signal & ZC Sequence (length = 16 order) \\
\hline
Multipath Delay Resolution & 50 ns \\
\hline
\end{tabular}
\label{tab:system_parameters}
\end{table}

\subsection{Measurement Procedure}
The Tx antenna was fixed on the rooftop of a 25 m-high residential building, while the Rx antenna was mounted on the rail of a vessel's deck as shown in Fig.~\ref{fig2}(c). A Kestrel 5500 weather meter was installed on the vessel to record meteorological parameters such as the wind speed, air temperature, humidity, and atmospheric pressure, as displayed in Fig.~\ref{fig2}(d). Additional marine environmental information, including sea water temperature and wave height, was recorded using the monitoring buoys in the sea area of Qidong City, as illustrated in Fig.~\ref{fig2}(e). The measurements were carried out over two consecutive days. On the first day, the average relative wind speed over the sea surface was about 7.7 m/s, higher than the average relative wind speed of 5.6 m/s on the second day, while the other measurement conditions were almost the same. Correspondingly, there were significant differences in the sea wave conditions between the two days.

On both days, the vessel carrying the Rx sailed away from the shore along the due east direction, and fixed-point measurements were performed at more than ten predetermined Rx locations as delineated in Fig. \ref{fig3}, with an Rx antenna height of 4 m. On the return trip of the vessel, measurements were taken without fixed Rx locations, maintaining a constant vessel speed of 9.6 knots while continuously recording the received signal strength. The vessel's direction was always aligned with the direction of the Tx antenna boresight, and the maximum Tx-Rx separation distance reached 33 km. 
On the second day, due to the decreased wind speed, additional fixed-point measurements with a lower Rx antenna height of 1 m were conducted for comparison. In the continuous-distance measurements, 14 sets of IQ data were collected within about 60 milliseconds at each measurement location, with adjacent measurement locations spaced about 0.02 km apart. While in the fixed-point measurements, 31 groups of the aforementioned 14 sets of IQ data were recorded with an approximately three-second time interval between adjacent groups, producing a total of 434 sets of IQ data at each location. More details of the measurement procedure are provided in Table~\ref{tab:measurement_summary}. As the measurements were conducted in winter, the sea surface temperature was around 3$^\circ C$, resulting in very weak evaporation effect~\cite{Wang18Access}. Therefore, the evaporation duct effect is neglected in the following analysis.

\begin{table*}[ht]
	\centering
	\caption{\textsc{Measurement Summary}}
	\begin{tabular}{c|cc}
	\hline
		\textbf{Date}                     & \textbf{Day 1}                          & \textbf{Day 2}                          \\ 
		\hline\hline
		\multirow{2}{*}{\textbf{Type}}    & {Fixed-Point (High Rx)}& {Fixed-Point (High and Low Rx)} \\ 		& {Continuous-Distance (High Rx)} & {Continuous-Distance (High Rx)} \\  
		\hline
		\multirow{2}{*}{\textbf{Tx-Rx Separation Distance (km)}} & Fixed-Point: 2.4, 3, 4, 5, 6, 7, 8, 9, 12, 15, 18, 21  & Fixed-Point: 3, 4, 5, 6, 7, 8, 9, 12, 15, 18, 21    \\
		& Continuous-Distance: 2.0 $\sim$ 24.6            & Continuous-Distance: 2.0 $\sim$ 33.8                   \\ 
		\hline
		{\textbf{Distance interval (km)}} & \multicolumn{2}{c}{Fixed-Point: 1 or 3; \quad Continuous-Distance: 0.02 } \\
		\hline
		{\textbf{Number of PDPs per Measurement Location}} & \multicolumn{2}{c}{Fixed-Point: 434; \quad Continuous-Distance: 14 } \\
		\hline
			\multirow{2}{*}{\textbf{Total Number of PDPs}} & Fixed-Point: 5208  & Fixed-Point: 4774 (High Rx), 4340 (Low Rx) \\
		& Continuous-Distance: 4984 & Continuous-Distance: 6888           \\ 
		\hline
		\textbf{Wind Speed (m/s)}         & About 7.7                                    & About 5.6                                     \\ 
		\hline
		\textbf{Sea Temperature (°C)}     & 3 $\sim$ 5                                    & 3 $\sim$ 4                                     \\ 
		\hline
		\textbf{Wave Height (m)}          & 0.5 $\sim$ 1.0                                & 0.4 $\sim$ 0.8                                 \\ 
		\hline
	\end{tabular}
	\label{tab:measurement_summary}
\end{table*}

\subsection{Data Processing}
\begin{figure}[!t]
	\centering
	\includegraphics[width=0.8\columnwidth]{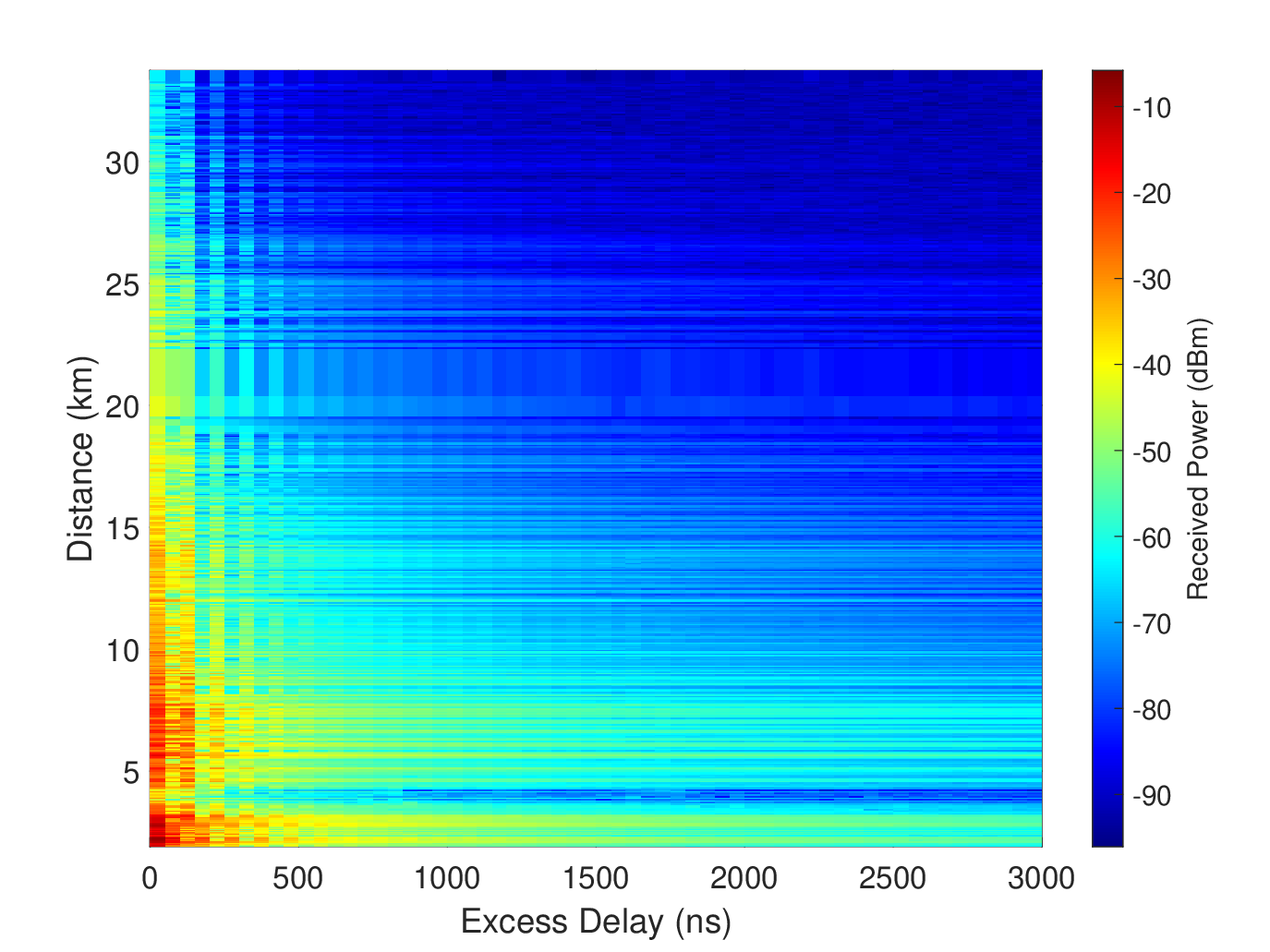}
	\caption{Examples of measured PDPs in the continuous-distance measurements on the second day.}
	\label{fig_PDP}
\end{figure}

The empirical CIR of the land-to-ship channel at 5.8 GHz was obtained by correlating the received IQ demodulated signal $y(t)$ with the transmitted ZC sequence $x(t)$. Since the autocorrelation of the ZC sequence can well approximate the delta function, the extracted CIR $\hat{h}(t, \tau)$ is given by (neglecting the noise temporarily for ease of exposition)
\begin{equation}
	\hat{h}(t, \tau)=x(t)\otimes y(t)=x(t)\otimes x(t)*h(t, \tau)=h(t, \tau),
\end{equation}

\noindent where $\otimes$ and $*$ denote the correlation and convolution operations, respectively, $h(t, \tau)$ is the ground-truth CIR.
The PDP is the squared-magnitude of the CIR, i.e., $|\hat{h}(t, \tau)|^2$. Examples of measured PDPs in the continuous-distance measurements are portrayed in Fig.~\ref{fig_PDP}.
\footnote{It is noteworthy that the received power drops substantially around 4 km compared with its adjacent distances in Fig.~\ref{fig_PDP}, corresponding to the most dominant peak in path loss in Fig.~\ref{fig4}(b), which is discussed in Section III. }

\textcolor{black}{A comprehensive characterization of maritime channels requires an understanding of different fading mechanisms occurring across distinct temporal scales. The instantaneous received power can be conceptually expressed as
\begin{equation}
P_{\text{Rx}} = P_{\text{Tx}}-\text{PL}-X_{\text{SWIFT}}-X_{\text{small-scale}},
\end{equation}
where $P_{\text{Rx}}$ represents the instantaneous received power in dBm, $P_{\text{Tx}}$ is the transmitted power in dBm, $\text{PL}$ stands for the distance-dependent large-scale path loss in dB, $X_{\text{SWIFT}}$ denotes the medium-scale fading in dB specifically induced by sea-wave fluctuations occurring on a temporal scale of seconds to tens of seconds, and $X_{\text{small-scale}}$ captures rapid short-term (millisecond-scale) variations in dB caused by multipath propagation. This formulation provides the framework for the comprehensive channel fading characterization in subsequent sections.}

\section{Large-Scale Fading}\label{sec_PL}
In this section, we analyze large-scale fading in maritime channels leveraging empirical path loss models and measured data.

\subsection{Empirical Distance-Dependent Path Loss Models}
In free space, the electromagnetic wave propagation loss can be described using the free space (FS) path loss model, which is given by
\begin{equation}
    \text{PL}^\text{FS}(f,d)[\text{dB}] = 20\log_{10}\left(\frac{4\pi fd}{c}\right),
\label{eq2}
\end{equation}

\noindent where $f$ denotes the carrier frequency, $d$ is the distance between the Tx and Rx, and $c$ is the speed of light. However, the LoS transmission distance is affected by the curvature of the Earth. The maximum LoS transmission distance is related to the antenna heights and can be expressed as\cite{Rappaport-WirelessCommunicationsPrinciplesandPractice}
\begin{equation}
    d_{\text{LoS}} = \sqrt{h_\text{t}^2 + 2h_\text{t}r_\text{e}} + \sqrt{h_\text{r}^2 + 2h_\text{r}r_\text{e}},
\label{eq3}
\end{equation} 

\noindent where $r_\text{e}$ denotes the Earth radius, $h_\text{t}$ and $h_\text{r}$ are the heights of the Tx and Rx antennas, respectively. In maritime communications, since the reflection effect of the sea surface is significant, the reflected ray should also be considered in the path loss model. \textcolor{black}{For vertically polarized and grazing incidence waves, the reflection coefficient is often approximated as -1 in terrestrial communications} \cite{Rappaport-WirelessCommunicationsPrinciplesandPractice}. Consequently, we can derive a simplified expression for the path loss via a two-ray model as follows\textcolor{black}{, assuming $d\gg h_\text{t},h_\text{r}$} \cite{Rappaport-WirelessCommunicationsPrinciplesandPractice}
\begin{equation}    
    \text{PL}^\text{two-ray}[\text{dB}] = 20\log_{10}\left[\frac{4\pi d}{\lambda}\bigg/\left|2\sin\left(\frac{2\pi h_\text{t}h_\text{r}}{\lambda d}\right)\right|\right],
\label{eq4}
\end{equation}

\noindent where $\lambda$ is the wavelength. The sine function in \eqref{eq4} causes the two-ray model to generate multiple peaks within a threshold $d_{\text{break}}$, which is identified as the point of occurrence of the last maximum of \(\sin\left(\frac{2\pi h_\text{t}h_\text{r}}{\lambda d}\right)\)\cite{Rappaport-WirelessCommunicationsPrinciplesandPractice} and is given by $d_{\text{break}} = \frac{4h_\text{t}h_\text{r}}{\lambda}.$

\textcolor{black}{In realistic maritime channels, unlike the traditional two-ray model, which assumes an idealized reflection coefficient of -1, we develop a modified two-ray (MTR) model influenced by multiple physical factors as follows}
\begin{equation}
    \text{PL}^\text{MTR}[\text{dB}]=20\log_{10}\left[\frac{4\pi d}{\lambda\left|1+ D S R \Gamma \exp\left(-j\frac{2\pi\Delta d}{\lambda }\right)\right|}\right],
    \label{eq_mod}
\end{equation}
where $\Delta d$ is the distance difference between the direct ray and reflected ray, $\Gamma$ is the reflection coefficient of the sea surface (set as -1 in \eqref{eq4}). The divergence factor $D$ accounts for the ray bundle spreading due to the Earth's curvature and is calculated as \cite{Parsons2000}
\begin{equation}
D = \left[1+\frac{2d_1d_2}{r_e(h_\text{t}+h_\text{r})}\right]^{-\frac{1}{2}},
\end{equation}
where $d_1$ and $d_2$ are the horizontal distances from the Tx and Rx to the reflection point, respectively.
The fixed-point shadowing factor $S$ in \eqref{eq_mod} considers the electromagnetic shielding effects of sea waves under grazing incidence conditions \cite{Smith1967TAP}
\begin{equation}
S = \frac{1-\frac{1}{2}\text{erfc}\left(\frac{\tan\theta_i}{\sqrt{2}{\beta_0}}\right)}{\Lambda(\tan\theta_i)+1},
\end{equation}
where
\begin{equation}
\Lambda(\tan\theta_i) = \frac{1}{2}\Bigg\{\sqrt{\frac{2}{\pi}}\frac{\beta_0 \exp[{-\left(\frac{\tan\theta_i}{\sqrt{2}\beta_0}\right)^2}]}{\tan\theta_i}- \text{erfc}\left(\frac{\tan\theta_i}{\sqrt{2}\beta_0}\right)\Bigg\}.
\label{eq_Lambda}
\end{equation}
In~\eqref{eq_Lambda}, $\theta_i$ is the incident angle, $\text{erfc}$ denotes the complementary error function, $\beta_0$ is the RMS slope of the sea surface with typical values determined by the empirical formula \cite{Cox1954}: $\beta_0=0.003+0.00512v_\text{w}$, with $v_\text{w}$ representing the wind speed in m/s. The sea surface roughness factor $R$ in \eqref{eq_mod} is calculated using the Miller-Brown model \cite{Miller1984}
\begin{equation}
R = \exp\left[\frac{-(4\pi\sigma_s\sin\theta_i)^2}{2\lambda^2}\right] I_0\left[\frac{-(4\pi\sigma_s\sin\theta_i)^2}{2\lambda^2}\right],
\end{equation}
where $I_0(\cdot)$ is the zero-order modified Bessel function of the first kind, $\sigma_s$ is the standard deviation of sea surface height with $\sigma_s = 0.0051v_\text{w}^2$.

Furthermore, additional diffraction loss for long-distance transmission needs to be considered. The curvature of the Earth's surface not only obstructs the LoS transmission but also blocks the reflected path from the sea surface. When the clearance in the first Fresnel zone is less than 60\% of the Fresnel zone radius, additional diffraction loss should be incorporated into the overall path loss prediction. This clearance distance can be denoted as \cite{ITU-RP.1546}
\begin{equation}
    d_\text{0.6F} = \frac{0.00015949 \cdot f \cdot h_\text{t} \cdot h_\text{r} \cdot (\sqrt{h_\text{t}} + \sqrt{h_\text{r}})}{0.0000389 \cdot f \cdot h_\text{t} \cdot h_\text{r} + 4.1 \cdot (\sqrt{h_\text{t}} + \sqrt{h_\text{r}})}.
\label{eq6}  
\end{equation}

\begin{figure}[!t]
    \centering
    \begin{subfigure}[t]{\columnwidth}
        \centering
        \includegraphics[width=0.9\columnwidth]{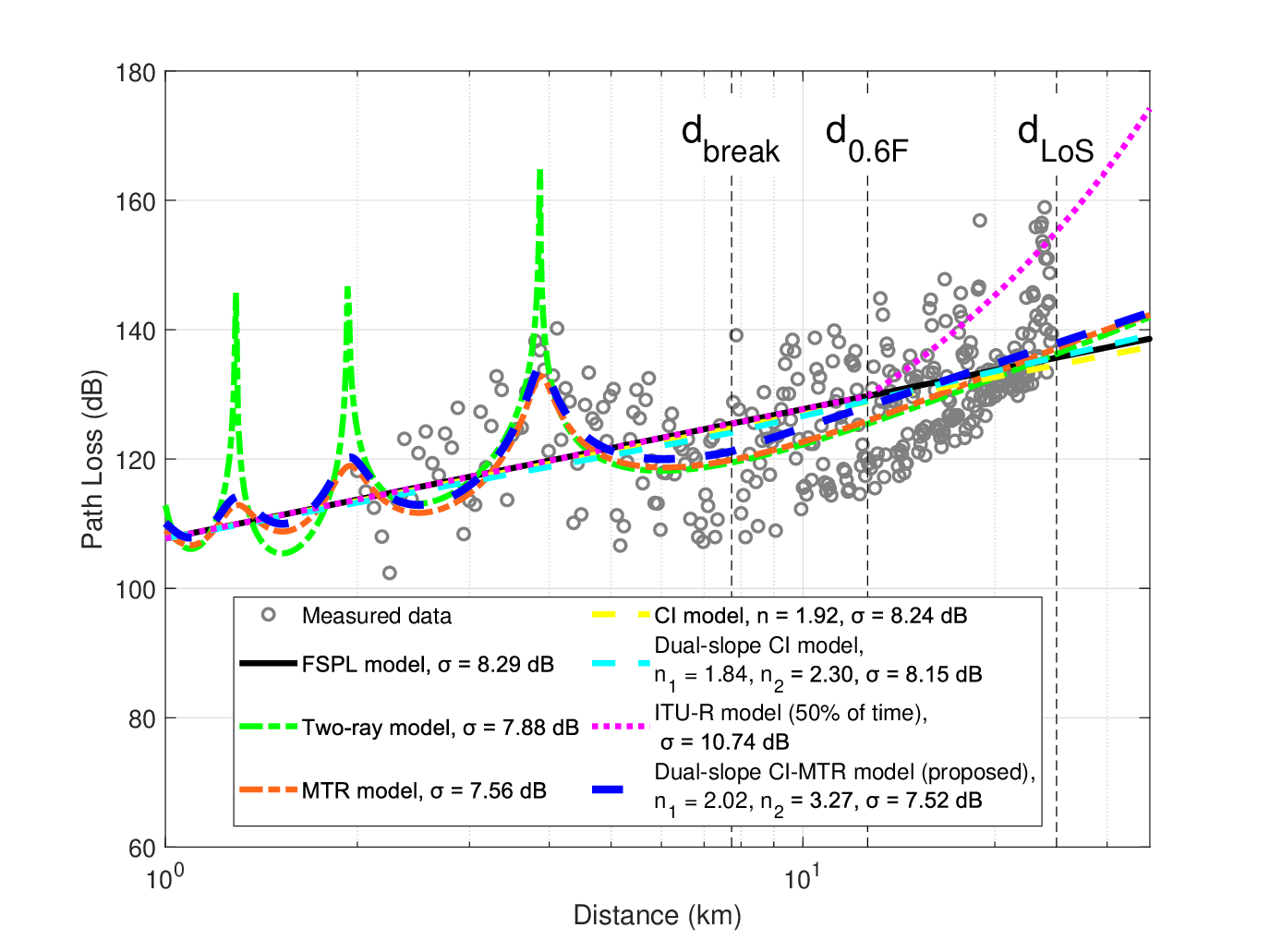}
        \caption{Day 1}
        
    \end{subfigure}
    \begin{subfigure}[t]{\columnwidth}
        \centering
        \includegraphics[width=0.9\columnwidth]{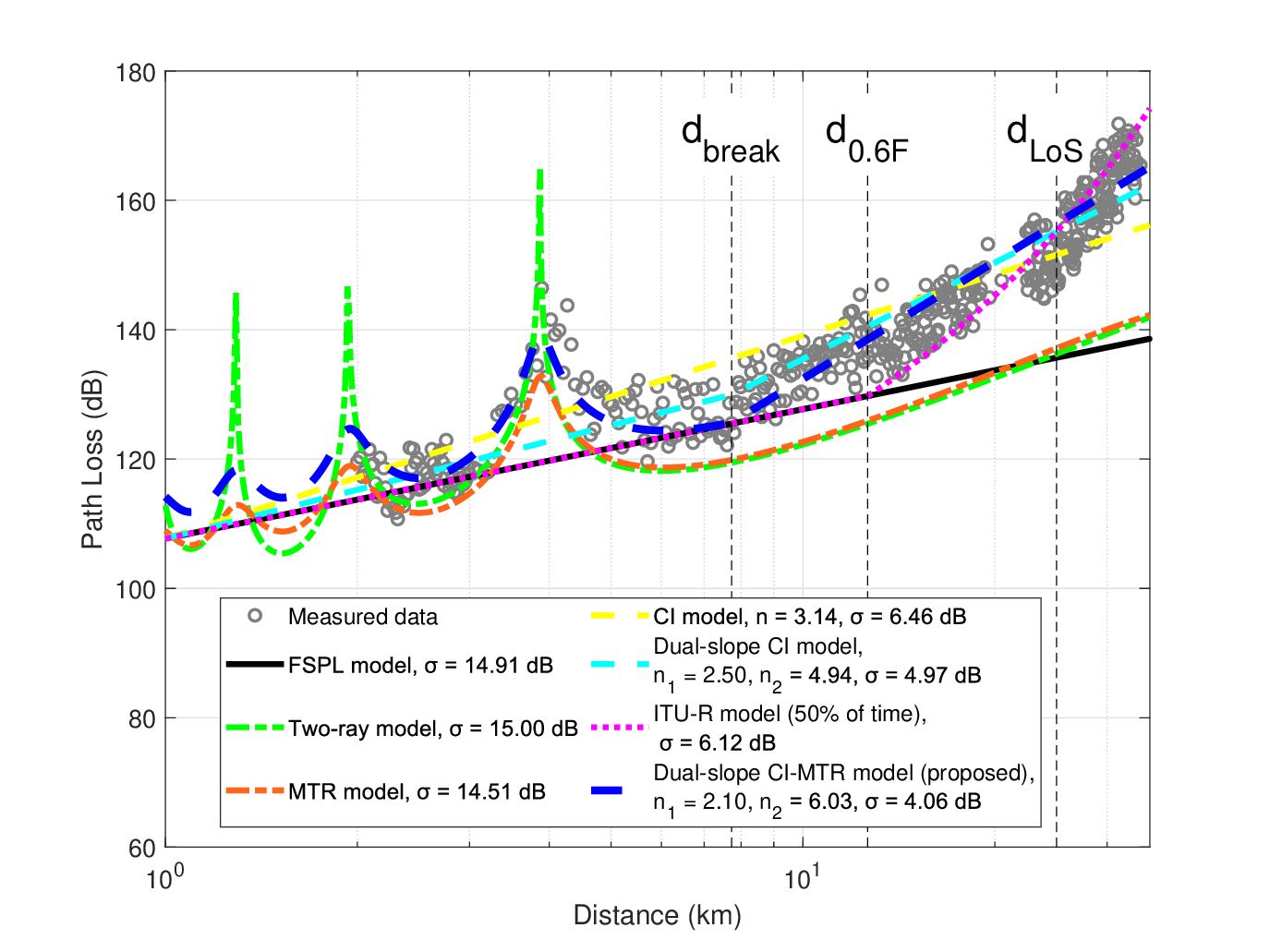}
        \caption{Day 2}
        
    \end{subfigure}
    \caption{\textcolor{black}{Measured data and various path loss models for the two measurement days, with parameters and performance metrics summarized in Table III.}}
    \label{fig4}
\end{figure}

\arrayrulecolor{black}
\color{black} 

In addition to the two-ray model, the close-in (CI) free space reference distance model~\cite{Sun16TVT} and the dual-slope CI model~\cite{Haneda16ICC} are also representative and widely used path loss models, where the CI model is an optional path loss model in the 3GPP standards~\cite{3GPP} as well. The dual-slope CI model takes $d_{\text{break}}$ as a threshold distance, aiming to better describe the peak characteristics within $d_{\text{break}}$ and fading changes beyond it. These two models are expressed as in~\eqref{eq_CI} and~\eqref{eq_dualCI}, respectively~\cite{Sun16TVT,Haneda16ICC}
\begin{equation}
	\text{PL}^\text{CI}(f,d)[\text{dB}] = \text{PL}^\text{FS}(f, d_0)[\text{dB}] + 10n \log_{10}\left(\frac{d}{d_0}\right)+X_\sigma^\text{CI}.
	\label{eq_CI}
\end{equation}
\begin{figure*}[h!]
\begin{equation}
\text{PL}^{\text{Dual-CI}}(f,d)[\text{dB}]\\ = 
	\begin{cases} 
		\text{PL}^\text{FS}(f,d_0) + 10n_1 \log_{10}\left(\frac{d}{d_0}\right),\quad\quad\quad\quad\quad\quad\quad\quad\quad~~\text{for } d_0<d \leq d_{\text{break}} \\ 
		\text{PL}^\text{FS}(f,d_0) + 10n_1 \log_{10}\left(\frac{d_\text{break}}{d_0}\right) + 10n_2 \log_{10}\left(\frac{d}{d_\text{break}}\right),\text{ }\text{for } d > d_{\text{break}} 
	\end{cases}
+X_\sigma^\text{Dual-CI}
	\label{eq_dualCI}
\end{equation}
\hrule
\end{figure*}
	
\noindent In~\eqref{eq_CI} and~\eqref{eq_dualCI}, $\text{PL}^\text{CI}$ and $\text{PL}^\text{Dual-CI}$ denote the path loss in dB over frequency and distance, $n$, $n_1$, and $n_2$ represent the path loss exponents (PLEs) within different distance ranges, and $X_\sigma^\text{CI}$ and $X_\sigma^{\text{Dual-CI}}$ are zero-mean log-normally distributed random variables describing large-scale shadow fading (SF). 


\subsection{Measured Results and Proposed Path Loss Model}

\begin{figure*}[h!]
	\begin{equation}
        \textcolor{black}{
		\text{PL}^{\text{Dual-CI-MTR}}(f,d)[\text{dB}]\\ = 
		\begin{cases} 
			10n_1\log_{10}\left[\frac{4\pi fd}{c\left|1+ D S R\Gamma \exp\left(-j\frac{2\pi f \Delta d}{c }\right)\right|}\right], \quad\quad\quad\quad\quad\quad\quad\quad~\text{for } d\leq d_{\text{break}} \\ 
			10n_1\log_{10}\left[\frac{4\pi fd_\text{break}}{c\left|1+ D S R\Gamma \exp\left(-j\frac{2\pi f \Delta d}{c }\right)\right|}\right] + 10n_2 \log_{10}\left(\frac{d}{d_\text{break}}\right),\text{ }\text{for } d > d_{\text{break}} 
		\end{cases}
	+X_\sigma^\text{Dual-CI-MTR}
		\label{eq_dualCITR}
        }
	\end{equation}
\end{figure*}

Measured path loss data are shown in Fig. \ref{fig4} for both measurement days, along with several empirical path loss models with fitted parameters and the aforementioned threshold distances ($d_{\text{break}}$, $d_\text{0.6F}$, $d_\text{LoS}$). The path loss model according to the ITU-R P.1812-7\cite{ITU-RP.1812-7} standard for 50\% of the time is also shown as a reference. \textcolor{black}{The fitting parameters in all the path loss models are derived by minimizing the RMSE between the model and measured data~\cite{Rappaport-WirelessCommunicationsPrinciplesandPractice,Sun16TVT}. The fitted parameter values and RMSEs for all the path loss models are listed in Table \ref{table5}.} As unveiled by Fig. \ref{fig4}, the measured path loss on the second day exhibits less fluctuation compared with the first day, with more pronounced peak characteristics within $d_{\text{break}}$, which aligns with the calmer sea conditions observed on the second day. Under the more windy conditions on the first day, the ideal single reflected path over the sea surface was disrupted, resulting in a more complex scattering environment, which is manifested in Fig. \ref{fig4} as lower measured path loss values at some positions on the first day in contrast to the second day.

\begin{table}
	\centering
	\caption{\textsc{\textcolor{black}{Path loss model comparison}}}
	\arrayrulecolor{black}  
	\color{black}           
	\begin{tabular}{|>{\centering\arraybackslash}p{1.2cm}|c|c|c|c|c|}
		\hline
		\multirow{2}{*}{\textbf{Model}} & \multirow{2}{*}{{\makecell{\textbf{No. of} \\ \textbf{Params}}}} & \multicolumn{2}{c|}{\textbf{PLE Values}} & \multicolumn{2}{c|}{\textbf{RMSE $\sigma$ (dB)}} \\
		\cline{3-6}
		& & \textbf{Day 1} & \textbf{Day 2} & \textbf{Day 1} & \textbf{Day 2} \\
		\hline
		FSPL & 0 & -- & -- & 8.29 & 14.91 \\
		\hline
		Two-ray & 0 & -- & -- & 7.88 & 15.00 \\
		\hline
		MTR & 0 & -- & -- & 7.56 & 14.51\\
		\hline
		CI & 1 & $n = 1.92$ & $n = 3.14$ & 8.24 & 6.46 \\
		\hline
		Dual-CI & 2 & \makecell{$n_1 = 1.84$ \\ $n_2 = 2.30$} & \makecell{$n_1 = 2.50$ \\ $n_2 = 4.94$} & 8.16 & 5.09 \\
		\hline
		ITU-R & 0 & -- & -- & 10.68 & 6.82 \\
		\hline
		Dual-slope CI-MTR & 2 & \makecell{$n_1 = 2.02$ \\ $n_2 = 3.27$} & \makecell{$n_1 = 2.10$ \\ $n_2 = 6.03$} & \cellcolor{gray!20}7.52 & \cellcolor{gray!20}4.06 \\
		\hline
	\end{tabular}
	\label{table5}
\end{table}

Moreover, as indicated by the RMSEs in Fig. \ref{fig4} and Table \ref{table5}, among the empirical path loss models introduced above, the \textcolor{black}{MTR} model matches the measured path loss the best for the first day. For the second day, however, the \textcolor{black}{MTR} model yields an unacceptably large fitting error, especially for distances beyond $d_\text{break}$. 
\textcolor{black}{
The MTR model incorporates realistic physical loss factors through the reflection coefficient, resulting in peaks within $d_{\text{break}}$ that better match the actual measurement data, while degenerating to the traditional two-ray model beyond $d_{\text{break}}$.}
On the other hand, the dual-slope CI model exhibits relatively high accuracy for both days in general, but it obviously fails to depict the fluctuation behavior of the path loss within $d_\text{break}$.

In view of the trend of the measured data and observations \textcolor{black}{in Fig. \ref{fig4}}, we propound a new path loss model named dual-slope \textcolor{black}{CI-MTR} model integrating the CI and \textcolor{black}{MTR} models, which is expressed in~\eqref{eq_dualCITR}. In this model, the \textcolor{black}{MTR} model incorporating an adaptive PLE $n_1$ is proposed to delineate the path loss within $d_\text{break}$, in conjunction with a second slope $n_2$ similar to that in the dual-slope CI model for distances beyond $d_\text{break}$. 
\textcolor{black}{As listed in Table \ref{table5}, the} slope values in the dual-slope CI-MTR model are similar to those in the dual-slope CI model in relevant literature (e.g., around 2 and 6 for calm sea in\cite{Joe07WCNC}), with lower $\sigma$ (e.g., over 8 dB in the literature) probably due to improved fitting within $d_\text{break}$. Since no standardized maritime path loss model with PLEs exists to our best knowledge, we compare the PLEs and SF standard deviations in our dual-slope CI-MTR model against 3GPP TR 38.901’s rural macro (RMa) LoS scenario. Most PLEs align (2 and 4 for the two slopes), except for $n_2$ on Day 2, likely due to Earth-curvature diffraction effects absent in the 3GPP RMa's 10 km range limit. Besides, the $\sigma$ on Day 1 is higher than the RMa scenario (4-6 dB), implying windy environments render more widely spread path loss values.

\textcolor{black}{\textit{Remark 1:} The dual-slope CI-MTR model captures the oscillating behavior of the path loss within $d_\text{break}$, as demonstrated by the multiple peaks of the dashed blue curve in Fig. \ref{fig4}. Furthermore, it allows for more flexible and accurate characterization of the path loss based on environmental conditions via $n_1$ and $n_2$.}

\section{SWIFT Fading}\label{sec_SWIF}
\begin{figure}[t]
	\centering
	\includegraphics[width=0.9\columnwidth]{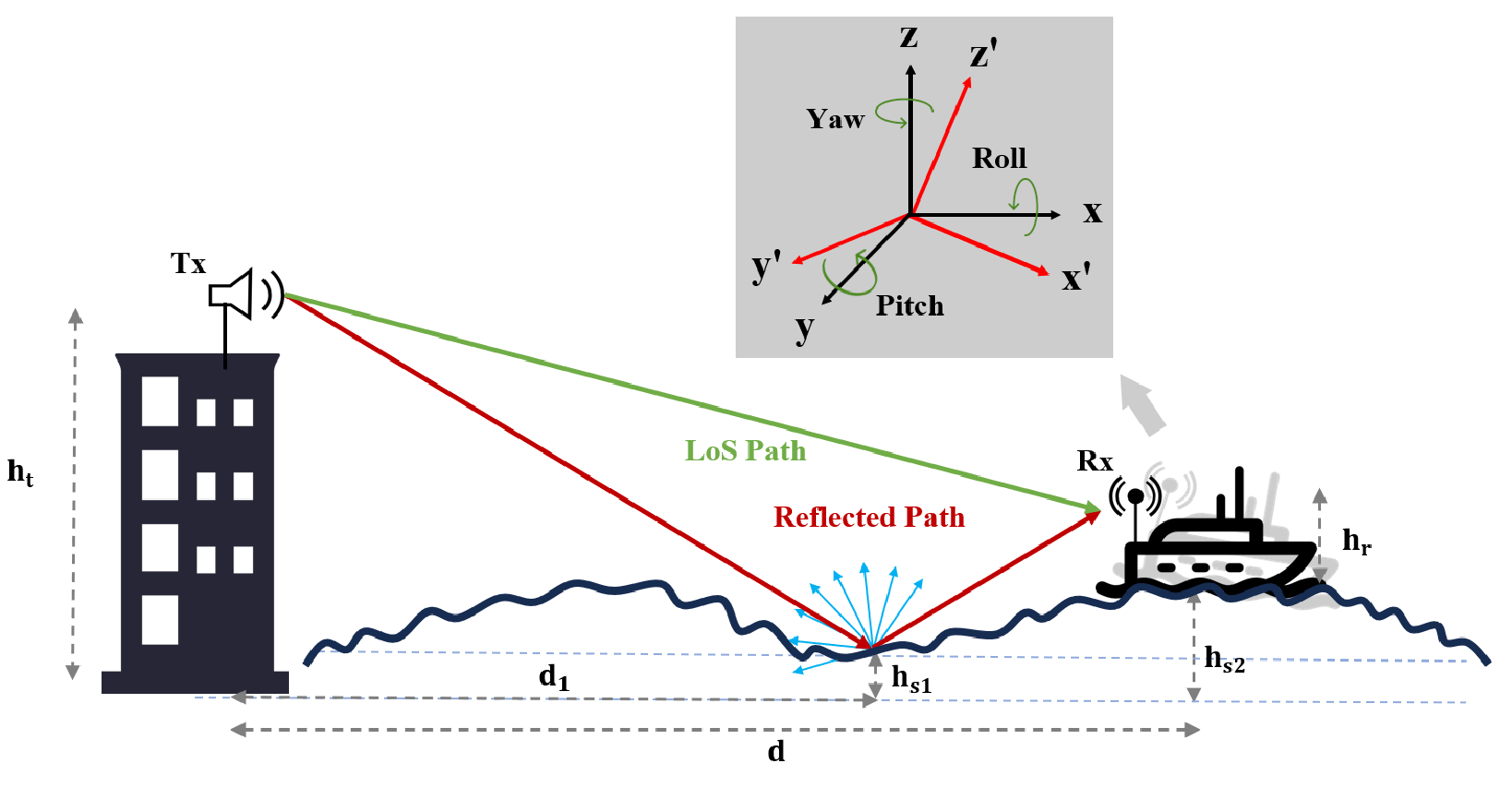}
	\caption{\textcolor{black}{Illustration of influencing factors on SWIFT fading (not drawn to scale), where $d_1$, $h_\text{s1}$, and $h_\text{s2}$ denote the horizontal distance between the Tx and the reflection point on the sea surface, the height of the reflection point relative to the calm sea surface, and the Rx antenna height relative to the calm sea surface subtracted by $h_\text{r}$, respectively.}}
	\label{fig10}
\end{figure}

\textcolor{black}{In maritime communications, sea wave fluctuations induce ship and antenna motions while altering reflected path amplitudes and phases, causing medium-timescale (wave-period dependent) signal strength variations, a phenomenon we term SWIFT fading. To model such fading, we incorporate both sea surface height variations that change Tx and Rx antenna relative heights, as well as antenna rotational motion causing pattern and polarization mismatch losses.}
As shown in Fig. \ref{fig10}, the height $h_\text{s}$ of fluctuating sea surface can be simplified with composite simple harmonic waves as~\cite{Pierson76rep}
\begin{equation}
	h_\text{s}(t,d) = \sum_i^{\infty} A_i\sin{\left(\frac{2\pi}{T_i}t-\frac{2\pi}{\lambda_i}d+\xi_i\right)},
	\label{eq:hs}
\end{equation}

\noindent where $T_i$ and $\lambda_i$ are the fluctuation period and wavelength of the $i$-th composite wave, respectively, $\xi_i$ denotes the initial phase of the $i$-th composite wave that follows a uniform random distribution in $[0,2\pi)$. $A_i$ represents the amplitude of the $i$-th composite wave, which depends on the frequency spectrum of sea waves $S(\omega)$, and the classic Pierson-Moskowitz (P-M) spectrum is often utilized to model $S(\omega)$ for low computational complexity~\cite{He22TCommun}
\begin{equation}
A_i = \sqrt{2S(\omega_i)\Delta \omega}, \quad S_\text{P-M}(\omega) = \frac{a_0g^2}{\omega^5}\exp[-\beta\left(\frac{g}{v_\text{w}\omega}\right)^4],
\end{equation}
where $\omega_i$ is the angular frequency with $\omega_i=2\pi/T_i$, $\Delta \omega$ denotes the frequency interval between adjacent harmonic waves, $a_0 = 8.1 \times 10^{-3}$, $\beta = 0.74$, $g$ is the gravitational acceleration equal to $9.81~\text{m/s}^2$.

As illustrated in Fig.~\ref{fig10}, denoting by $h_\text{s1}$ and $h_\text{s2}$ the height changes of the reflection point and the Rx antenna with respect to the calm sea surface, respectively, due to sea wave fluctuation, then the equivalent Tx antenna height relative to the reflection surface becomes
\begin{equation}
	h_\text{t1}(t,d)=h_\text{t}-h_\text{s1}(t,d_1),
	\label{eq_ht1}
\end{equation}

\noindent where $d_1$ represents the horizontal distance between the Tx and the reflection point on the fluctuating sea surface, and is a function of $d$. Similarly, the equivalent Rx antenna height against the reflection surface is recast as
\begin{equation}
	h_\text{r1}(t,d)=h_\text{r}+h_\text{s2}(t,d)-h_\text{s1}(t,d_1).
	\label{eq_hr1}
\end{equation}

\noindent Based on the geometric relation, we have
\begin{equation}
	\frac{d_1}{d-d_1}=\frac{h_\text{t1}(t,d)}{h_\text{r1}(t,d)}.
	\label{eq_d1}
\end{equation}

\begin{figure*}
    \begin{equation}
    \textcolor{black}{
    \mathbf{R}_x(\phi)=
  \begin{bmatrix}
    1&0&0\\
    0&\cos\phi&-\sin\phi\\
    0&\sin\phi& \cos\phi
  \end{bmatrix},\quad
  \mathbf{R}_y(\theta)=
  \begin{bmatrix}
    \cos\theta&0&\sin\theta\\
    0&1&0\\
    -\sin\theta&0&\cos\theta
  \end{bmatrix},\quad
  \mathbf{R}_z(\psi)=
  \begin{bmatrix}
    \cos\psi&-\sin\psi&0\\
    \sin\psi& \cos\psi&0\\
    0&0&1
  \end{bmatrix}.
    }
    \label{eq_matrix}
    \end{equation}
\end{figure*}

\noindent For given values of $d$ and time instants, the corresponding $d_1$ and $h_\text{s1}$ can be acquired by solving \eqref{eq_d1} numerically. 


\textcolor{black}{For the rotational motion of the Rx antenna, also as shown in Fig.~\ref{fig10}, we define three types of rotational motion, i.e., roll, pitch, and yaw, around the x, y, and z coordinate axes, respectively. With the Rx antenna center as the origin and rotation center, and the vessel's forward direction as the x-axis direction, these three rotational motions can be described by the rotation matrices in (\ref{eq_matrix}), where we assume that the roll angel $\phi$, pitch angle $\theta$, and yaw angle $\psi$ follow sinusoidal motions as $\phi(t)=\Phi\sin(\omega_\phi t + \alpha_\phi),
    \theta(t)=\Theta\sin(\omega_\theta t + \alpha_\theta),
    \psi(t)=\Psi\sin(\omega_\psi t + \alpha_\psi)$
    where $\Phi$, $\Theta$, and $\Psi$ are the maximum rotation angles, $\omega_\phi$, $\omega_\theta$, and $\omega_\psi$ are the frequencies of the rotation angle variations, $\alpha_\phi$, $\alpha_\theta$, and $\alpha_\psi$ follow a uniform random distribution in $[0, 2\pi]$.}

\textcolor{black}{Considering that the shore-to-ship direction aligns with the x-axis, the unit LoS vector $\mathbf{u}_0$ is $[\cos\alpha_0, 0, \sin\alpha_0]$, where $\alpha_0 = \arctan\frac{h_t-h_r}{d}$. The direction vector of the Rx antenna after rotation is $\mathbf{u} = \mathbf{R}_z(\psi)\mathbf{R}_y(\theta)\mathbf{R}_x(\phi)\mathbf{u}_0 = [u_x, u_y, u_z]$. For an omnidirectional antenna, we only consider the impact of elevation angle changes on antenna gain. The Rx antenna elevation angle relative to the horizontal plane is $\arcsin(u_z)$, while the antenna gain loss due to antenna pattern rotation is $L_\text{g} = 20\log_{10}(F(\arcsin(u_z)))$, where $F(\cdot)$ is the radiation pattern of the omnidirectional antenna, and $u_z = -\sin(\theta)\cos(\alpha_0) + \cos(\theta)\cos(\phi)\sin(\alpha_0)$. For polarization mismatch loss $L_p$, since both Tx and Rx antennas are vertically polarized, we consider the inner product of polarization vectors as the polarization mismatch loss. Defining the Tx and Rx polarization vectors $\mathbf{p}_t$ and $\mathbf{p}_r$ as $[0,0,1]^T$, the inner product of polarization vectors after rotation is $\rho = |\mathbf{p}_t^T\mathbf{R}_z(\psi)\mathbf{R}_y(\theta)\mathbf{R}_x(\phi)\mathbf{p}_r| = |\cos(\theta)\cos(\phi)|$, and $L_\text{p} = 20\log_{10}(\rho)$.
}

\begin{figure}[t]
	\centering
	\includegraphics[width=0.75\columnwidth]{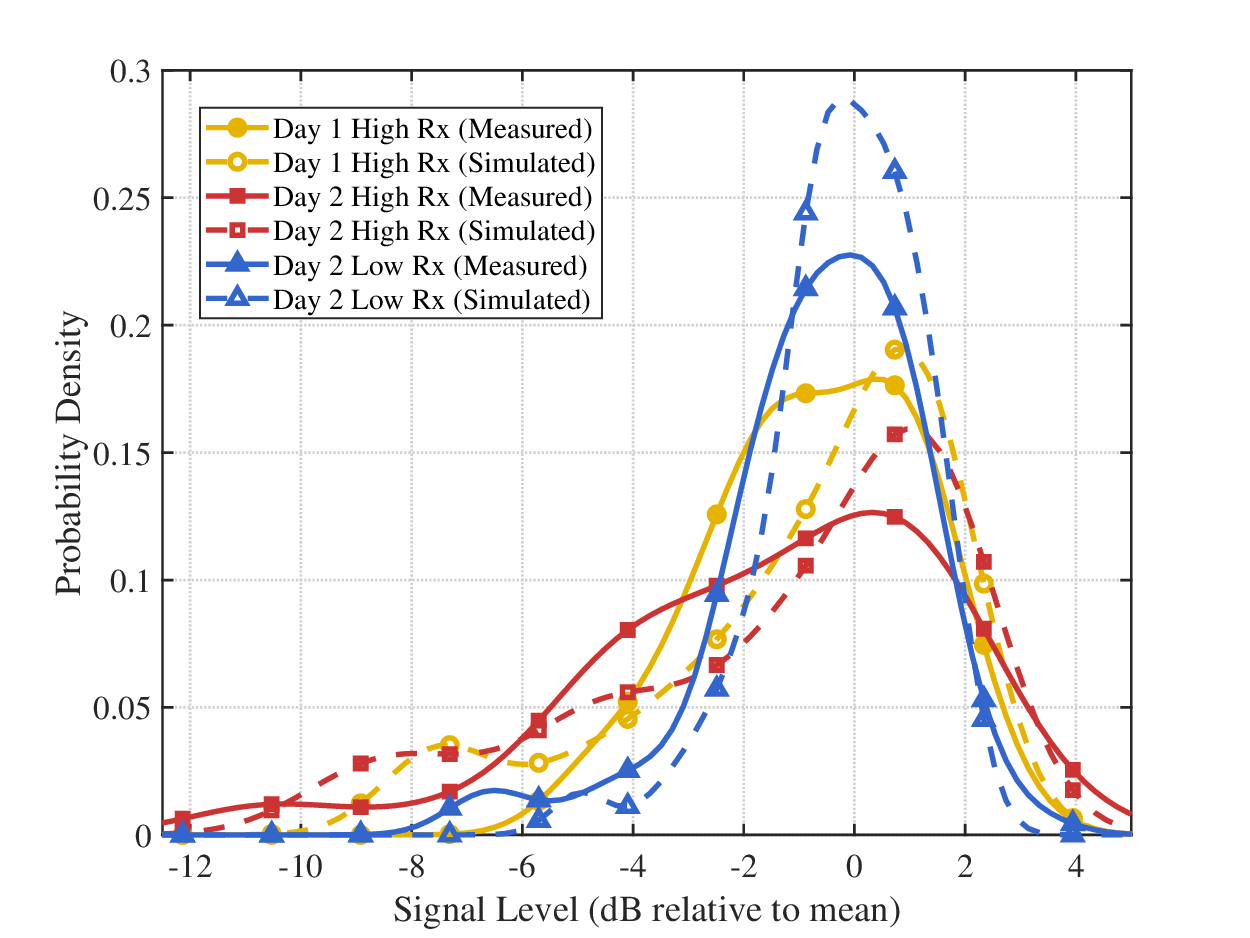}
	\caption{\textcolor{black}{Measured data and simulated SWIFT fading based on the MTR model with sea wave fluctuations and antenna motions incorporated.}}
	\label{fig11}
\end{figure}

\textcolor{black}{After substituting $h_\text{t}$ and $h_\text{r}$ in \eqref{eq_mod} with $h_\text{t1}$ in \eqref{eq_ht1} and $h_\text{r1}$ in \eqref{eq_hr1}, respectively, and incorporating the losses $L_\text{g}$ and $L_\text{p}$, we obtain a new model capable of simulating SWIFT fading.} In Fig. \ref{fig11}, we compare the \textcolor{black}{probability density functions (PDFs)} of SWIFT fading extracted from the measured and simulated data respectively under different conditions\footnote{\textcolor{black}{In our simulations, five harmonic
waves with equally spaced frequencies and randomly generated initial phases are utilized in~\eqref{eq:hs}, which provides an effective balance between computational complexity
 and the accuracy required to capture realistic sea-wave-induced antenna height variations and
subsequent SWIFT fading.}}. The signal levels are calculated by first averaging the small-scale fading across the 14 PDPs within each 60-ms time interval, resulting in 31 received signal voltage amplitudes at each fixed location, and then computing the deviations from their mean value at each location. 

\textit{Remark 2:} As can be observed from Fig. \ref{fig11}, \textcolor{black}{both measured and simulated signal variations are} more severe for high Rx antennas, attributed to the fact that the amplitude of swaying and fluctuation is more acute for a higher Rx antenna. The proposed SWIFT fading model can well capture the fading intensity changing trend as the antenna height and sea states vary. On the other hand, the simulated signal strength is generally more concentrated than the measured ones, since \eqref{eq:hs} and the vessel motion models may be simplified compared with the actual circumstances.

\textcolor{black}{It is noteworthy that the performance of the proposed SWIFT fading model might be limited under extreme sea conditions or for larger antenna displacements, as these scenarios could introduce additional propagation dynamics not fully captured by the current formulation.}


\section{Small-Scale Fading}\label{sec_small-scale}
\begin{figure*}[t]
	\centering
	\begin{subfigure}[b]{0.33\textwidth}
		\centering
		\includegraphics[width=0.95\textwidth]{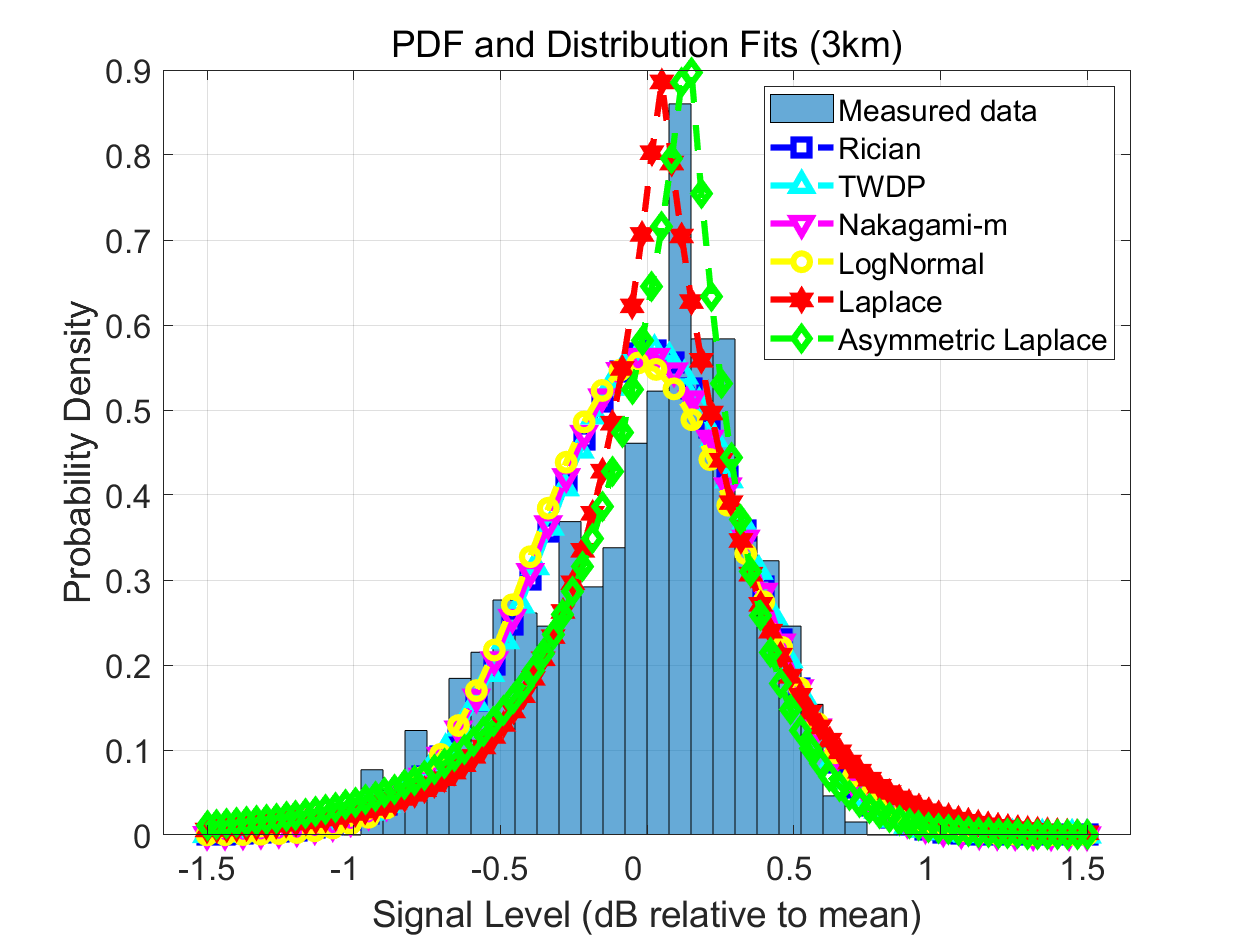}
		\caption{Day 1 high Rx}
		\label{fig:day1}
	\end{subfigure}\hfill
	\begin{subfigure}[b]{0.33\textwidth}
		\centering
		\includegraphics[width=0.95\textwidth]{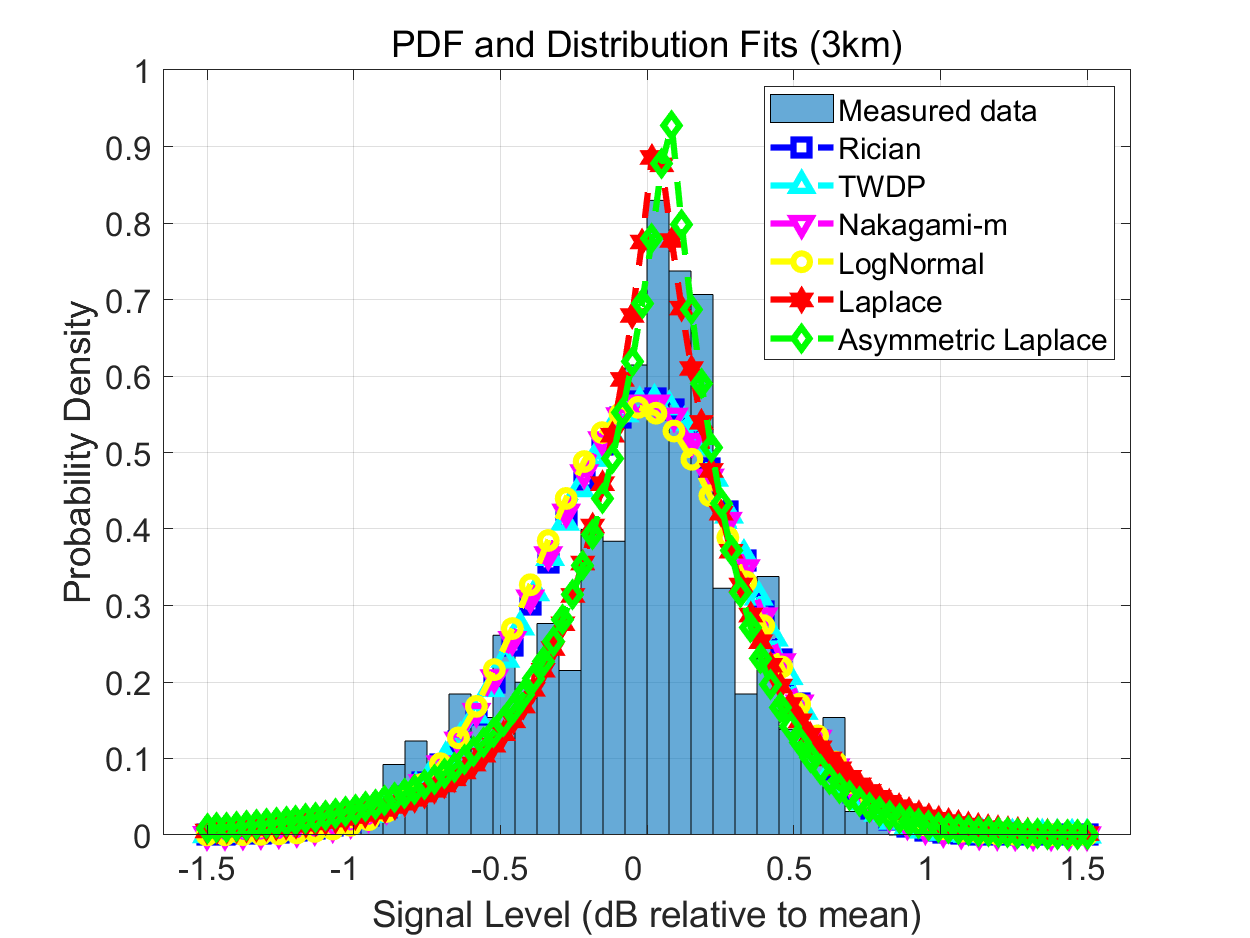}
		\caption{Day 2 high Rx}
		\label{fig:day2}
	\end{subfigure}\hfill
	\begin{subfigure}[b]{0.33\textwidth}
		\centering
		\includegraphics[width=0.95\textwidth]{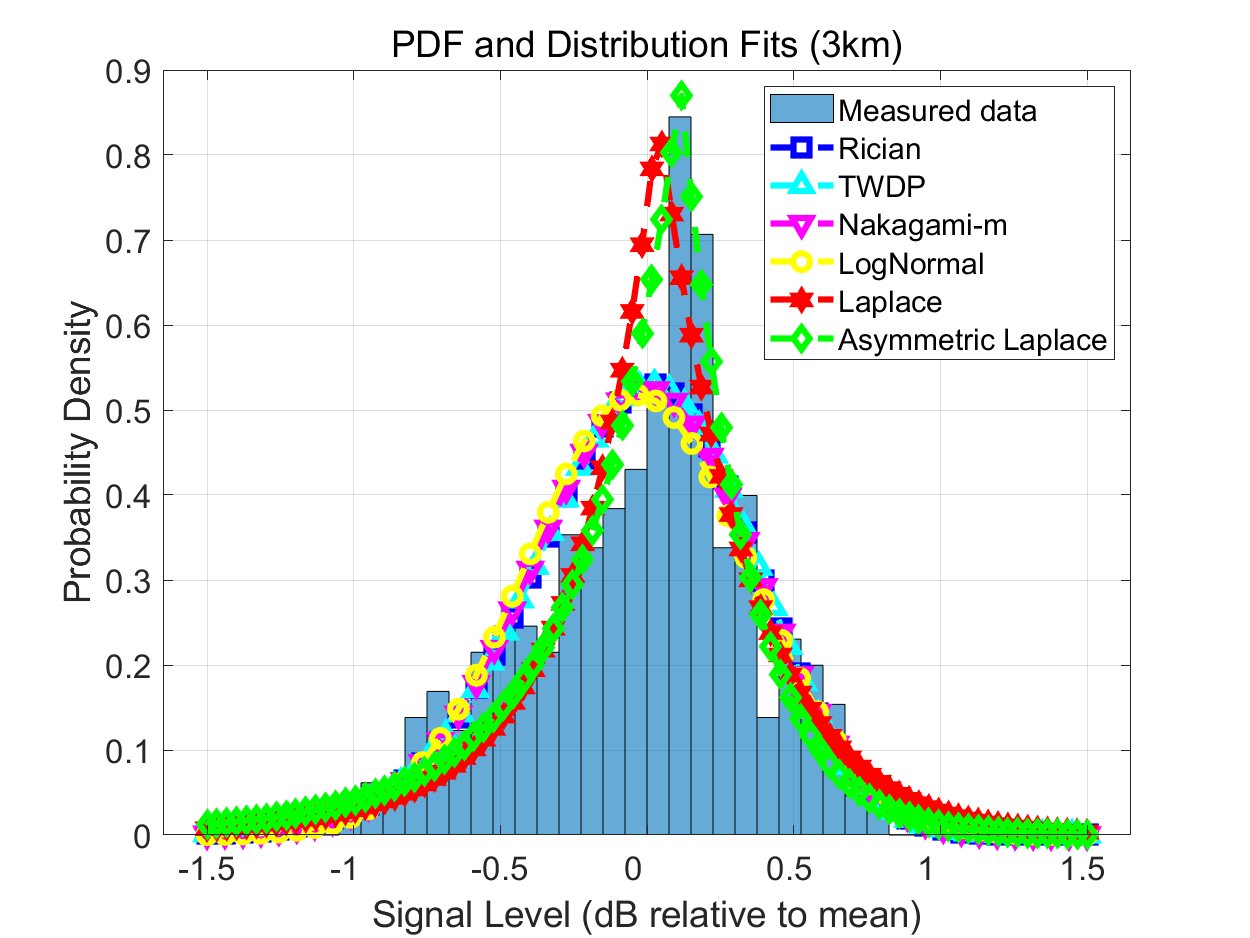}
		\caption{Day 2 low Rx}
		\label{fig:day2low}
	\end{subfigure}
	\caption{Measured data and empirical models for small-scale fading at 3 km.}
	\label{fig:3km}
\end{figure*}

\begin{figure*}[t]
	\centering
	\begin{subfigure}[b]{0.33\textwidth}
		\centering
		\includegraphics[width=0.95\textwidth]{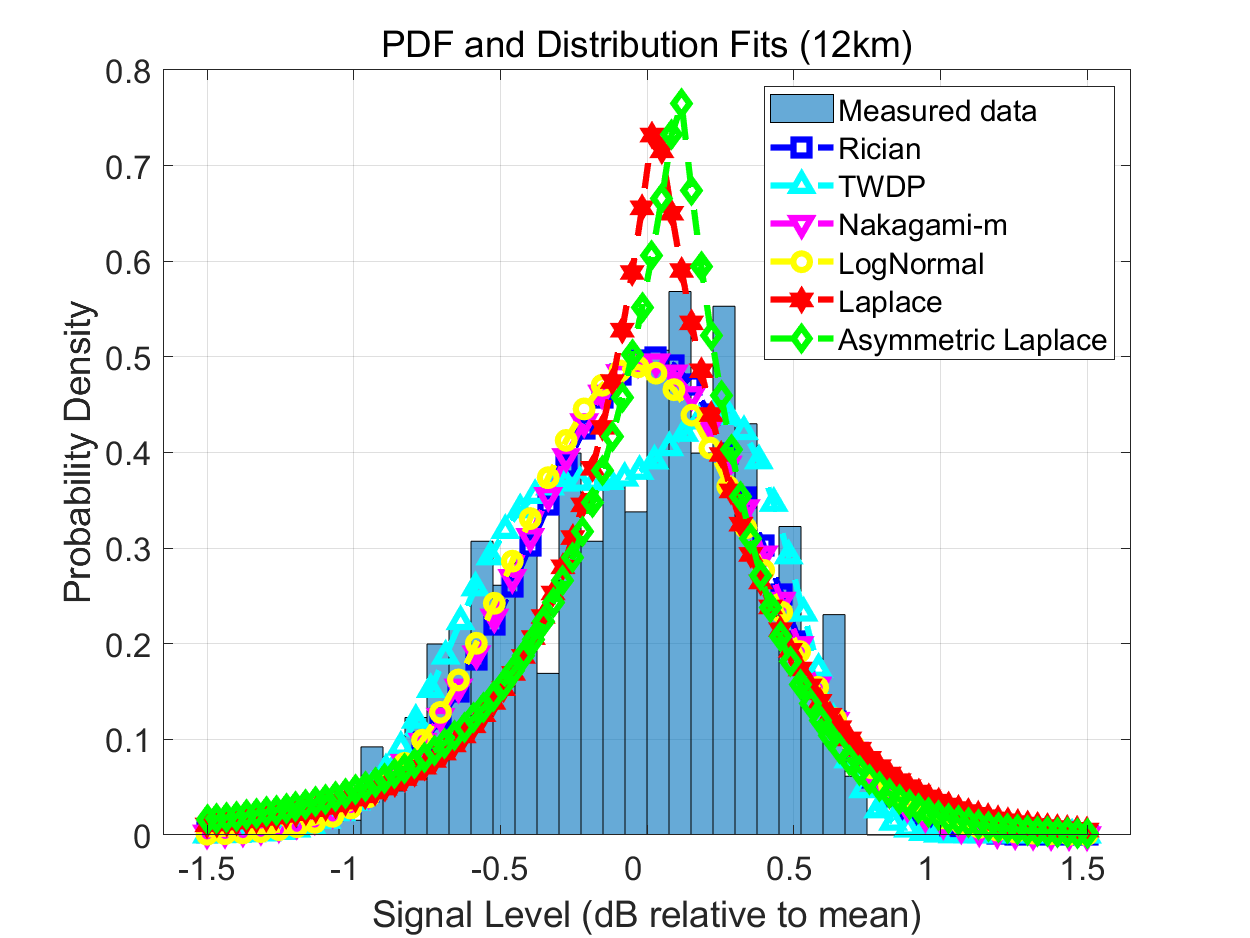}
		\caption{Day 1 high Rx}
		\label{fig:day1}
	\end{subfigure}\hfill
	\begin{subfigure}[b]{0.33\textwidth}
		\centering
		\includegraphics[width=0.95\textwidth]{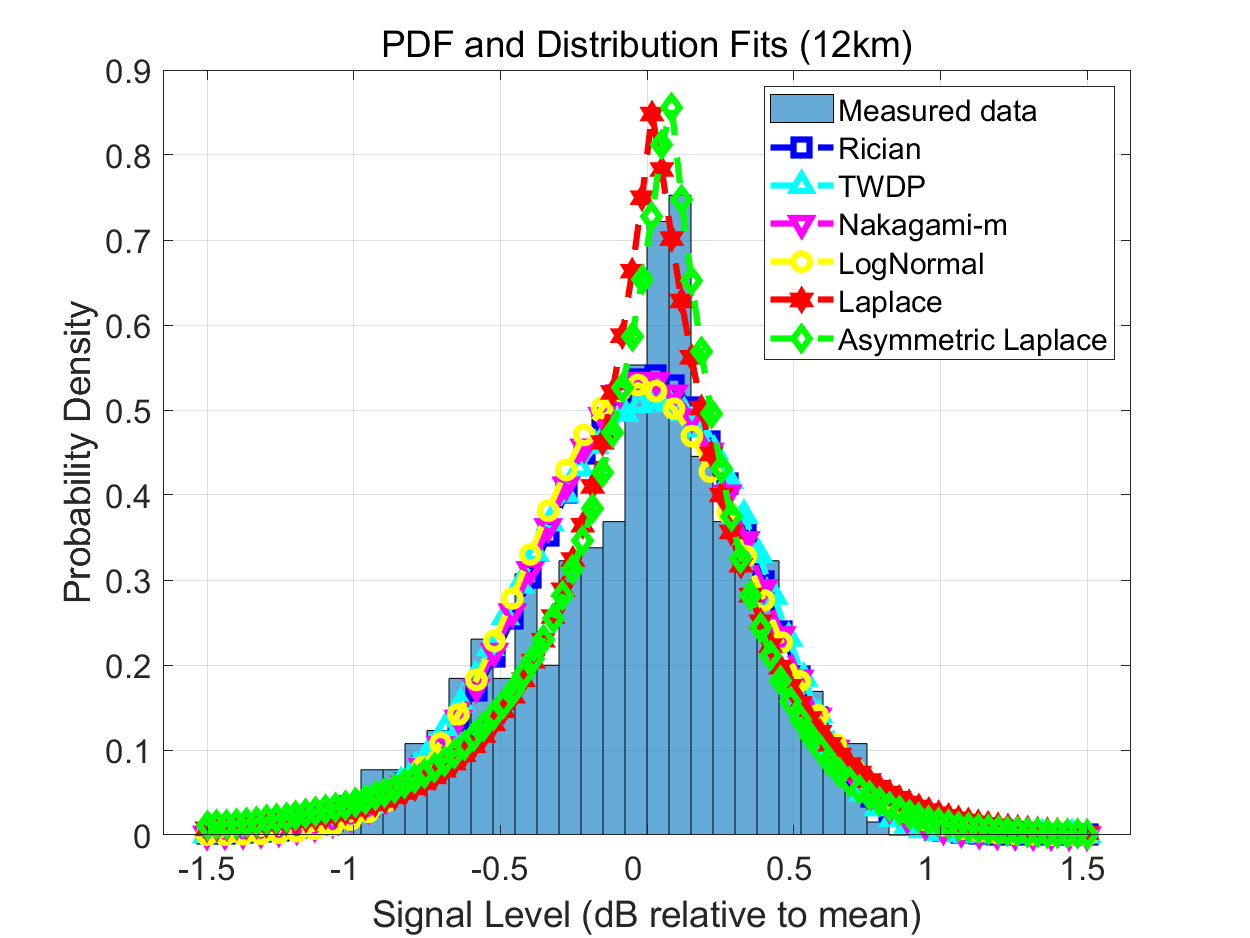}
		\caption{Day 2 high Rx}
		\label{fig:day2}
	\end{subfigure}\hfill
	\begin{subfigure}[b]{0.33\textwidth}
		\centering
		\includegraphics[width=0.95\textwidth]{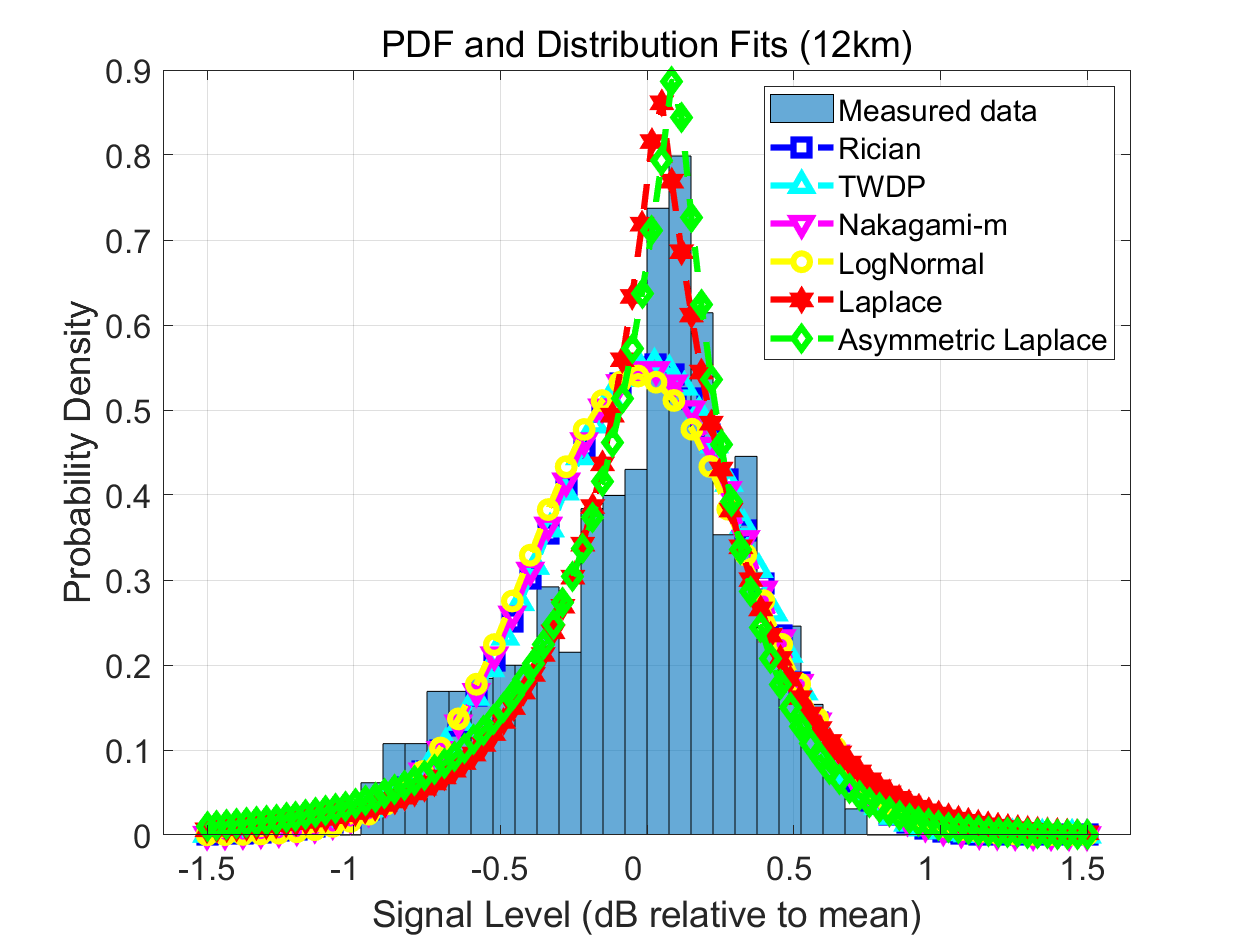}
		\caption{Day 2 low Rx}
		\label{fig:day2low}
	\end{subfigure}
	\caption{Measured data and empirical models for small-scale fading at 12 km.}
	\label{fig:12km}
\end{figure*}

As mentioned previously, for the fixed-point measurements, 31 sets of PDPs were collected per location at three-second intervals. To examine the small-scale deviations in the received signal voltage amplitude, we calculate the deviations relative to the mean value for each set to remove the impacts of large-scale fading and SWIFT fading. The small-scale fading results for different days (i.e., environmental conditions), Rx heights, and distances are illustrated in Figs. \ref{fig:3km} and \ref{fig:12km}.

In marine environments within \textcolor{black}{$d_{\text{LoS}}$}, propagation is dominated by direct paths and sea-surface reflections due to sparse obstacles. Additionally, the rough sea surface creates multiple scattered paths, effectively acting as fragmented mirrors. These conditions make the TWDP model suitable within LoS distances, particularly under \textcolor{black}{adverse sea conditions}~\cite{Wang18Access}. The mathematical expression of the TWDP model is \cite{Durgin02ITC}
\begin{equation}
	\tilde{V} = V_1 \exp(j\psi_1) + V_2 \exp(j\psi_2) + \tilde{V}_{\text{dif}},
\end{equation}

\noindent \textcolor{black}{where \(\tilde{V}\) is the received baseband signal, \(V_1\) and \(V_2\) (0 $<$ \(V_2\) $<$ \(V_1\)) represent the constant amplitudes of two dominant specular multipath components, with their phases \(\psi_1\) and \(\psi_2\) independently and uniformly distributed over the range $[0,2\pi)$. The diffuse component $\tilde{V}_{\text{dif}}$ is modeled as a complex Gaussian random variable with zero mean and variance $2\sigma^2$, representing the combined contribution of numerous weaker multipath waves.}
 In the case where \(V_2=0\), the TWDP model simplifies to the Rician distribution. When both \(V_1\) and \(V_2\) are zero, the TWDP model further degenerates to the Rayleigh distribution. The \textcolor{black}{PDF} of the TWDP model is given by~\cite{Durgin02ITC}
\begin{equation}
	\begin{aligned}
		f_U(u) &=  \frac{u}{\pi \sigma^2}\exp\left(-\frac{u^2}{2\sigma^2}-B_1\right)\\
        &\cdot \int_0^\pi \exp\left(B_1 B_2 \cos x\right) I_0 \left[ \frac{u\sqrt{2B_1(1 - \Delta \cos x)}}{\sigma}  \right] dx ,
	\end{aligned}
\end{equation}
where $U$ denotes the envelope of $\tilde{V}$, $B_1=\frac{V_1^2 + V_2^2}{2\sigma^2}$, $B_2=\frac{2V_1V_2}{V_1^2 + V_2^2}$, and the diffuse component power is $2\sigma^2$.

Moreover, Figs. \ref{fig:3km} and \ref{fig:12km} show a pronounced spike near 0 dB of mean value\footnote{We have also tried the median value instead of the mean, and have obtained similar trends.} (slightly right-skewed) in most measurements, indicating concentrated received voltage amplitudes with occasional deep fades. To obtain proper statistical distributions of the small-scale fading, we consider the TWDP model along with several other stochastic distributions commonly used to describe small-scale fading, including the \textcolor{black}{Rician, Nakagami-m, and lognormal distributions. For better fitting of the central peak, we also exploit the Laplace and asymmetric Laplace distributions whose expressions are given by}
\textcolor{black}{
\begin{equation}
		f_{\text{Laplace}}(x) = \frac{1}{2b} \exp\left(-\frac{|x-\mu|}{b}\right),
\end{equation}
\begin{equation}
		f_{\text{Asymmetric Laplace}}(x) = 
		\frac{1}{b_1 + b_2} 
		\begin{cases} 
			\exp\left(\frac{x - \mu}{b_1}\right) & \text{if } x < \mu \\
			\exp\left(-\frac{x - \mu}{b_2}\right) & \text{if } x \geq \mu 
		\end{cases},
\end{equation}
}
	
\noindent \textcolor{black}{where $\mu$ is the location parameter, while $b$, $b_1$, and $b_2$ are scale parameters. The Laplace distribution is well-suited for fitting data concentrated around the mean and have many extreme values. The asymmetric Laplace distribution, on the other hand, allows for different scale parameters on either side of the mean, introducing skewness into the distribution. This flexibility enables it to more effectively fit data with asymmetrical tail behavior.} 

\begin{table*}[t]
	\centering
	\caption{\textsc{Fitted Parameters and Evaluation Metrics for Small-Scale Fading at 3 km}}\label{tbl_exist}
	\begin{tabular}{|>{\centering\arraybackslash}c|c|c|c|c|c|c|c|c|c|}
		\hline
		\multirow{2}{*}{\textbf{Distribution}} & \multicolumn{3}{c|}{\textbf{Day1 \textcolor{black}{High Rx}}} & \multicolumn{3}{c|}{\textbf{Day2 \textcolor{black}{High Rx}}}  & \multicolumn{3}{c|}{\textbf{Day2 \textcolor{black}{Low Rx}}}\\ 
		\cline{2-10}
		& \textbf{Fitted Parameters} & \textbf{K-S}&\textbf{RMSE}& \textbf{Fitted Parameters} & \textbf{K-S}&\textbf{RMSE}& \textbf{Fitted Parameters} & \textbf{K-S}&\textbf{RMSE}\\ 
		\hline  
		{\makecell* {\textbf{Rician} \\$f(x|s,\sigma)$}} & {\makecell* {$K=18.806 \text{ dB}$\\$s=0.994$\\$\sigma=0.081$}}  & 0.074 & 0.847  & {\makecell* {$K=18.817 \text{ dB}$\\$s=0.994$\\$\sigma=0.081$}} & 0.078 & 0.877& {\makecell* {$K=18.153 \text{ dB}$\\$s=0.992$\\$\sigma=0.087$}}&0.062 &0.870\\
		\hline 
		{\makecell* {\textbf{TWDP} \\$f(x|K,\Delta,\sigma)$}} & {\makecell* {$K=18.804 \text{ dB}$\\$\Delta=0.004 $\\$\sigma=0.081$}} & 0.074 & 0.847 & {\makecell* {$K=18.817 \text{ dB}$\\$\Delta=0.008 $\\$\sigma=0.081$}}  & 0.078 & 0.877& {\makecell* {$K=18.154 \text{ dB}$\\$\Delta=0.001 $\\$\sigma=0.087$}}&0.062 &0.870\\
		\hline  
		{\makecell* {\textbf{Nakagami-m} \\$f(x|\mu,\omega)$}} & {\makecell* {$\mu=32.031$\\$\omega=1.015$}}  & 0.079 & 0.885 & {\makecell* {$\mu=38.219$\\$\omega=1.015$}} & 0.084 & 0.904&{\makecell* {$\mu=32.891$\\$\omega=1.017$}} &0.068 &0.897\\
		\hline 
		{\makecell* {\textbf{Lognormal} \\$f(x|\mu,\sigma)$}} & {\makecell* {$\mu=-0.007$\\$\sigma=0.083$}} & 0.088 & 0.964 & {\makecell* {$\mu=-0.007$\\$\sigma=0.082$}} & 0.095 &0.965 & {\makecell* {$\mu=-0.008$\\$\sigma=0.089$}}& 0.080&0.957\\
		\hline  
		{\makecell* {\textbf{Laplace} \\$f(x|\mu, b)$}} & {\makecell* {$\mu=1.011$\\$b=0.065$}} & 0.090 & 0.951 & {\makecell* {$\mu=1.007$\\$b=0.062$}} & 0.070 &0.542 & {\makecell* {$\mu=1.009$\\$b=0.069$}}& 0.086&0.769\\
		\hline 
		{\makecell* {\textbf{Asymmetric Laplace} \\$f(x|\mu, b_1, b_2)$}} & {\makecell* {$\mu=1.033$\\$b_1=0.045$\\$b_2=0.081$}} & \cellcolor{gray!20}0.055 & \cellcolor{gray!20}0.645 & {\makecell* {$\mu=1.018$\\$b_1=0.051$\\$b_2=0.072$}} & \cellcolor{gray!20}0.042 &\cellcolor{gray!20}0.458 & {\makecell* {$\mu=1.027$\\$b_1=0.052$\\$b_2=0.082$}}& \cellcolor{gray!20}0.056&\cellcolor{gray!20}0.557\\
		\hline
	\end{tabular}
	\label{table2}
\end{table*}

\begin{table*}[h]
	\centering
	\caption{\textsc{Fitted Parameters and Evaluation Metrics for Small-Scale Fading at 12 km}}\label{tbl_exist}
	\begin{tabular}{|>{\centering\arraybackslash}c|c|c|c|c|c|c|c|c|c|}
		\hline
		\multirow{2}{*}{\textbf{Distribution}} & \multicolumn{3}{c|}{\textbf{Day1 \textcolor{black}{High Rx}}} & \multicolumn{3}{c|}{\textbf{Day2 \textcolor{black}{High Rx}}}  & \multicolumn{3}{c|}{\textbf{Day2 \textcolor{black}{Low Rx}}}\\ 
		\cline{2-10}
		& \textbf{Fitted Parameters} & \textbf{K-S}&\textbf{RMSE}& \textbf{Fitted Parameters} & \textbf{K-S}&\textbf{RMSE}& \textbf{Fitted Parameters} & \textbf{K-S}&\textbf{RMSE}\\ 
		\hline  
		{\makecell* {\textbf{Rician} \\$f(x|s,\sigma)$}} & {\makecell* {$K=17.640 \text{ dB}$\\$s=0.992$\\$\sigma=0.092$}}  & \cellcolor{gray!20}0.050 & 0.641  & {\makecell* {$K=18.328 \text{ dB}$\\$s=0.993$\\$\sigma=0.085$}} & 0.055 &0.799 & {\makecell* {$K=18.548 \text{ dB}$\\$s=0.993$\\$\sigma=0.083$}}&0.076 &0.872\\
		\hline 
		{\makecell* {\textbf{TWDP} \\$f(x|K,\Delta,\sigma)$}} & {\makecell* {$K=23.076 \text{ dB}$\\$\Delta=0.222 $\\$\sigma=0.049$}} & 0.053 & \cellcolor{gray!20}0.560& {\makecell* {$K=18.324 \text{ dB}$\\$\Delta=0.001 $\\$\sigma=0.085$}}  & 0.055 & 0.799& {\makecell* {$K=18.552 \text{ dB}$\\$\Delta=0.001 $\\$\sigma=0.083$}}&0.076 &0.872\\
		\hline  
		{\makecell* {\textbf{Nakagami-m} \\$f(x|\mu,\omega)$}} & {\makecell* {$\mu=29.319$\\$\omega=1.018$}}  & 0.055 & 0.661 & {\makecell* {$\mu=34.245$\\$\omega=1.016$}} & 0.061 & 0.825 &{\makecell* {$\mu=35.909$\\$\omega=1.016$}} & 0.082 & 0.908\\
		\hline 
		{\makecell* {\textbf{Lognormal} \\$f(x|\mu,\sigma)$}} & {\makecell* {$\mu=-0.009$\\$\sigma=0.094$}} & 0.067 & 0.711 & {\makecell* {$\mu=-0.007$\\$\sigma=0.087$}} & 0.072 & 0.884 & {\makecell* {$\mu=-0.007$\\$\sigma=0.085$}} & 0.092 & 0.984\\
		\hline  
		{\makecell* {\textbf{Laplace} \\$f(x|\mu, b)$}} & {\makecell* {$\mu=1.007$\\$b=0.076$}} & 0.092 & 0.917 & {\makecell* {$\mu=1.005$\\$b=0.067$}} & 0.067 &0.777 & {\makecell* {$\mu=1.009$\\$b=0.065$}}& 0.077&0.703\\
		\hline 
		{\makecell* {\textbf{Asymmetric Laplace} \\$f(x|\mu, b_1, b_2)$}} & {\makecell* {$\mu=1.026$\\$b_1=0.060$\\$b_2=0.090$}} & 0.070 & 0.903 & {\makecell* {$\mu=1.018$\\$b_1=0.056$\\$b_2=0.077$}} & \cellcolor{gray!20}0.046 &\cellcolor{gray!20}0.586 & {\makecell* {$\mu=1.023$\\$b_1=0.051$\\$b_2=0.077$}}& \cellcolor{gray!20}0.044&\cellcolor{gray!20}0.628\\
		\hline
	\end{tabular}
	\label{table3}
\end{table*}

Using maximum likelihood estimation, we fit the aforementioned distributions to measured small-scale fading data. The fitted parameters and evaluation metrics are presented in Tables \ref{table2} and \ref{table3} for Tx-Rx separation distances of 3 km and 12 km, respectively. The goodness of fit for each distribution is assessed using the K-S statistic and the RMSE. The former measures the maximum CDF difference, while the latter reflects the overall PDF discrepancy.
As illustrated in Fig. \ref{fig:3km} and Fig. \ref{fig:12km}, for the shorter 3-km link, the TWDP distribution degenerates to Rician, and the asymmetric Laplace distribution fits the best in this case. However, at a longer distance (12 km), the fitting results on the first day under a higher sea state show a notable increase in the small-scale diffuse component, consistent with more scattered paths from a larger rough sea surface at greater distances. The TWDP distribution best captures the envelope and shows bimodality of the small-scale fading. Its K-S statistic is slightly higher than that of the Rician distribution, probably due to insufficient data samples. On the second day, under a lower sea state, the TWDP distribution again degenerates to the Rician distribution \textcolor{black}{and the asymmetric Laplace distribution is still the best fit}. For longer distances, the weakening of the direct path and the reduced shadowing effect result in little differences between the small-scale fading distributions for high and low Rx heights.

\textit{Remark 3:} Tables \ref{table2} and \ref{table3} show that small-scale fading generally follows the asymmetric Laplace distribution, except for adverse sea conditions and long distances (e.g., 12 km), where the TWDP model better captures bimodal fading. This indicates that sea state conditions truly influence the small-scale fading characteristics of maritime wireless channels. Consequently, different small-scale fading models should be applied to depict the channel characteristics under varying sea state conditions.

\section{Channel Sparsity}\label{sec_sparsity}
In maritime communications, the sparse distribution of scatterers at sea incurs significant sparsity in the wireless channel\cite{Wang18Access}. While there exist a few potential metrics to evaluate channel sparsity, such as the number of multipath components (MPCs), channel degrees of freedom~\cite{Li25CC,Sun25CM}, Gini index\cite{Hurley09TIT}, and Rician $K$ factor~\cite{Zhang21TCCN}, the Gini index and Rician $K$ factor have been shown to be the most reasonable choices~\cite{Liu24TVT}. The Rician $K$ factor and Gini index $G$ are respectively defined as~\cite{Zhang21TCCN,Liu24TVT}
\begin{equation}
    K = \frac{P_{\text{LoS}}}{P_{\text{tot}}-P_{\text{LoS}}},
\end{equation}
\begin{equation}
    G = 1-2\sum_{n=1}^{N}\frac{P_{n}}{P_{\text{tot}}}\cdot\frac{N-n+1/2}{N},
\label{G}
\end{equation}

\noindent where $P_{\text{tot}}$ is the total power of all MPCs in a PDP, $P_{\text{LoS}}$ represents the power of the LoS or the strongest path, $N$ is the total number of MPCs in a PDP, and $P_n$ is the power of each MPC arranged in an ascending order, i.e., $P_1\leq P_2\leq\dots\leq P_N$. If the LoS path exists, we usually define that $P_{\text{LoS}}=P_N$, where the LoS path possesses the highest power among all MPCs. The Rician $K$ factor represents the ratio of the power of the LoS or the strongest path to the power of the remaining MPCs. A higher $K$ indicates greater channel sparsity. The Gini index measures power concentration across MPCs, which is increasingly being adopted as a standard for measuring channel sparsity~\cite{Liu24TVT}. As shown by (\ref{G}), the value of \( G \) ranges from 0 to 1. A larger \( G \) value indicates that power is concentrated in fewer MPCs, prompting a higher degree of sparsity. When \( G = 1 \), it suggests that all the power is concentrated in the LoS or strongest path, whereas \( G = 0 \) implies that the power is evenly distributed across all MPCs.\footnote{ \textcolor{black}{Note that channels exhibiting high sparsity in terms of the Gini
index and Rician $K$ factor imply that few MPCs contain most of the received power, naturally corresponding to a small number of significant MPCs.}} Unlike the Rician $K$ factor, the Gini index is scale-invariant and depends solely on power proportions, making it a more accurate measure of channel sparsity\cite{Hurley09TIT}.

Since the signal bandwidth (20 MHz) is relatively small in our measurements, leading to a relatively low multipath delay resolution, the number of resolvable MPCs may be smaller than it actually is. To obtain an estimation of the channel sparsity as accurate as possible, we first demonstrate that the Gini index acquired using~\eqref{G} can be considered as a lower bound of those corresponding to multipath delay resolutions equal to or finer than the one herein (i.e., 50 ns). When the multipath delay resolution is finer than 50 ns, there will be $M$ ($M\geq1$) resolvable MPCs within the time duration of each of the current $N$ resolvable MPCs. We consider the following two situations: 1) the special case where powers of the $M$ resolvable MPCs are identical, and 2) the general case where powers of the $M$ resolvable MPCs are not necessarily equal.

\textit{Lemma 1:} Denote with $\Delta\tau_1$ and $\Delta\tau_2$ the minimum resolvable multipath delay intervals corresponding to bandwidths of $B_1$ and $B_2$, respectively, where $B_2>B_1$ so that $\Delta\tau_2<\Delta\tau_1$. Denote the Gini index corresponding to $B_1$ as $G_1$. When the powers of the resolvable MPCs are equally distributed within each of the delay bins with width $\Delta\tau_1$, denote with $G_2$ the Gini index associated with the bandwidth $B_2$, then we have
\begin{equation}
	G_2=G_1.
\end{equation}

\textit{Proof:} Please see the appendix.

Lemma 1 states that when each of the MPCs in a PDP is split into a number of MPCs due to a finer multipath delay resolution, and the power is equally distributed among the newly-born resolvable MPCs, the Gini index remains unaltered, i.e., the channel sparsity level does not vary. In practice, nevertheless, the powers of the MPCs are usually not identical, corresponding to the general case mentioned above. Under this circumstance, we have the following lemma.

\textit{Lemma 2:} Denote with $G_3$ the Gini index in the general case where powers of the newly-born resolvable MPCs may differ from each other, then we obtain
\begin{equation}
	G_3\geq G_2,
\end{equation}

\noindent where $G_2$ represents the Gini index defined in Lemma 1.

\textit{Proof:} Please see Section V-A of~\cite{Liu24TVT}.

Lemma 2 implies that the channel sparsity is enhanced when the MPCs have variant powers. Combining Lemma 1 and Lemma 2, we have $G_3\geq G_1$, indicating that the Gini index and hence the channel sparsity tend to increase with the multipath delay resolution (or equivalently, the bandwidth) if all the other conditions remain unchanged. 

\begin{figure}[t]
	\centering
	\includegraphics[width=0.8\columnwidth]{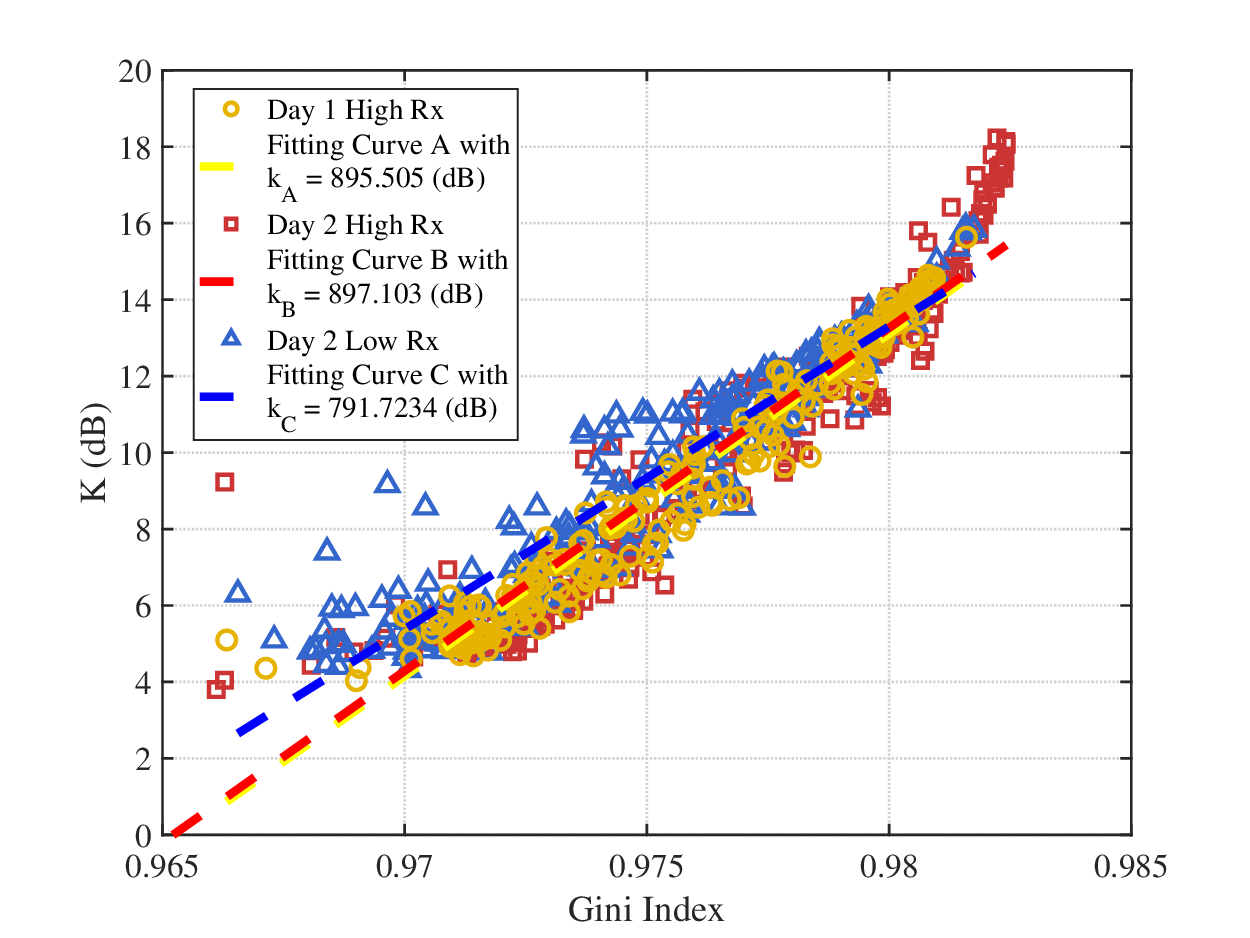}
	
	\caption{\textcolor{black}{Rician $K$ factor against Gini index for fixed-point measurements, where $\mathrm{k_A}$, $\mathrm{k_B}$, and $\mathrm{k_C}$ are slopes of the fitted curves.}}
	\label{fig9}
\end{figure}

Fig. \ref{fig9} illustrates the Rician $K$ factor versus the Gini index under different fixed-point measurement conditions across all distance ranges, along with least squares fits. \textcolor{black}{Note that we have examined variations of the Rician $K$ factor and the Gini index against distance, but have not observed any obvious trend. Thus, the results presented herein include all measured distances. It is evident from Fig. \ref{fig9} that the }maritime Gini index concentrates in the range of 0.965 to 0.985. As a comparison, the Gini index varies between about 0.4 and 0.8 at 5.9 GHz with a bandwidth of 30 MHz for the urban scenario~\cite{Zhang21TCCN}, and between about 0.75 and 0.92 at 6 GHz with a bandwidth of 400 MHz for the indoor office scenario~\cite{Liu24TVT}. According to Lemma 1 and Lemma 2, an increase in the bandwidth will raise the Gini index. Therefore, the maritime wireless channel is sparser than typical urban and indoor scenarios. More importantly, over all the fixed-point measurement data, the least squares fitting slope in Fig. \ref{fig9} suggests that a larger $K$ factor tends to be associated with a higher Gini index, which is consistent with the characteristics of increased channel sparsity. \textcolor{black}{Additionally, the Rician $K$ factor and the Gini index are slightly lower on Day 1 for the same antenna height, indicating that rougher sea conditions are likely to render more diffuse scattering power.}

\section{Temporal Statistics}\label{sec_temp}
\subsection{RMS Delay Spread}
\begin{figure}[t]
	\centering
	\includegraphics[width=0.8\columnwidth]{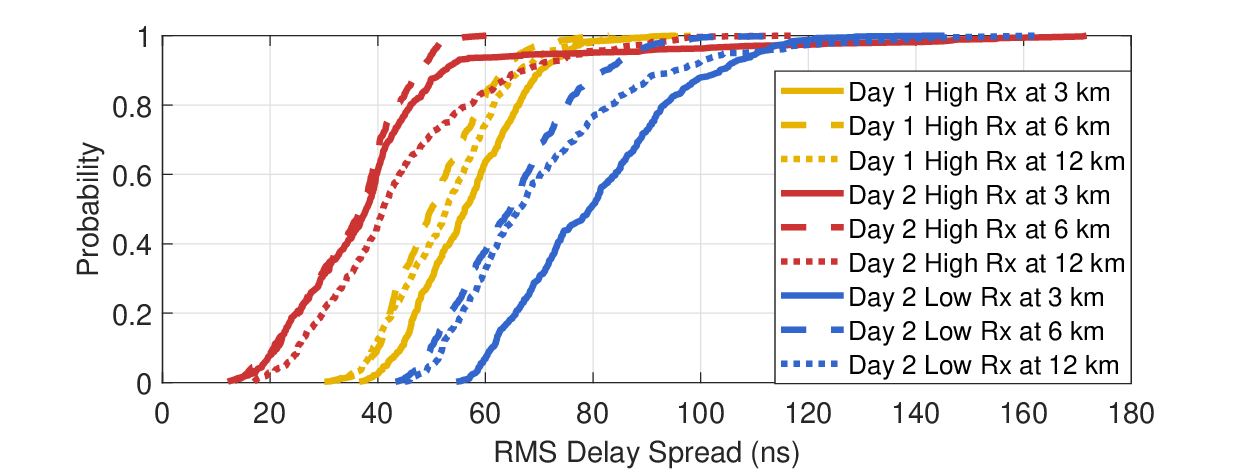}
	\caption{RMS delay spread for fixed-point measurements.}
	\label{fig5}
\end{figure}

RMS delay spread is a critical characteristic of a wireless propagation channel~\cite{Rappaport-WirelessCommunicationsPrinciplesandPractice}, which embodies the temporal dispersion of multipath signals. The RMS delay spread $\sigma_\tau$ is defined as
\textcolor{black}{
\begin{equation}
	\sigma_\tau = \sqrt{\frac{\sum_{n} \tau_n^2 P(\tau_n)}{\sum_{n} P(\tau_n)} - \left(\frac{\sum_{n} \tau_n P(\tau_n)}{\sum_{n} P(\tau_n)}\right)^2},
	\label{eq9}
\end{equation}
}

\noindent where \textcolor{black}{\( n \in [1, N] \) represents the index of MPCs, \(\tau_n\) denotes the delay of the $n$-th MPC, and \(P(\tau_n)\) is the received power corresponding to the delay \(\tau_n\)}.
The CDFs of RMS delay spread in the fixed-point measurements are shown in Fig. \ref{fig5}, which suggests that the distributions across different distances under the same condition do not vary significantly. For the high Rx, the RMS delay spread on the first day is concentrated between 40 ns and 80 ns, while on the second day, it ranges from about 20 ns to 60 ns, approximately 20 ns lower overall compared to the first day. This reduction is attributed to the calmer sea surface on the subsequent day, which decreases the variation range of scattered paths. Conversely, the RMS delay spread for the low Rx on the second day shows an increase, likely stemming from enhanced sea-surface scattering reception due to a lower antenna height. The variation in RMS delay spread necessitates the development of environment-adaptive maritime signal transmission schemes, so as to dynamically accommodate coherence bandwidth fluctuations and mitigate inter-symbol interference that can cause signal distortion.

\subsection{MPC Power-Delay Distribution}
In this subsection, we examine how the MPC power varies with respect to the excess time delay. Inspired by the measured PDPs and the standard channel model of the 3GPP~\cite{3GPP}, a single-slope exponential PDP can be assumed, where the mean power of the MPC decays exponentially with the excess delay and can be expressed as
\begin{equation}
P_n=\overline{P_0}\exp\left(-\frac{\tau_n}{\gamma}\right),
\end{equation}
	
\noindent where $P_n$ and $\tau_n$ denote the mean power and excess delay of the $n$-th resolvable MPC, respectively, $\overline{P_0}$ represents the average power of the first resolvable MPC, and $\gamma$ stands for the excess delay at which the average MPC power attenuates to $1/e$ of its original value. Note that both $P_n$ and $\overline{P_0}$ are normalized by the total received power in a PDP. 

\begin{figure}[!t]
	\centering
	\begin{subfigure}[b]{0.162\textwidth}
		\centering
		\includegraphics[width=\textwidth]{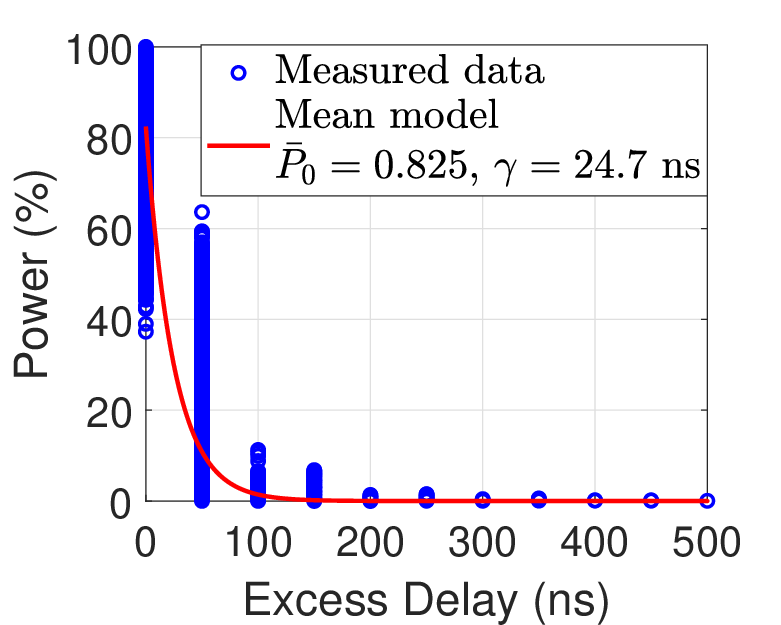}
		\caption{Day 1 High Rx}
		\label{fig:day1}
	\end{subfigure}%
	\hfill
	\begin{subfigure}[b]{0.162\textwidth}
		\centering
		\includegraphics[width=\textwidth]{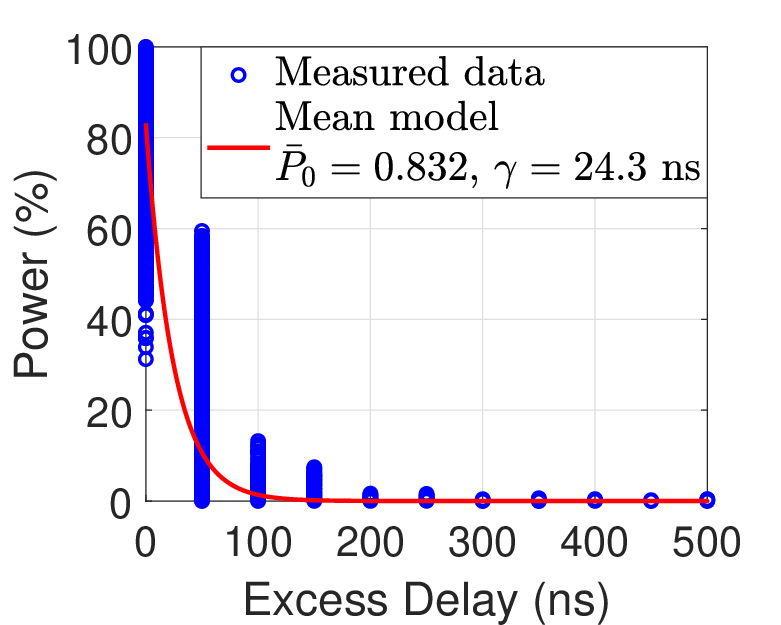}
		\caption{Day 2 High Rx}
		\label{fig:day2}
	\end{subfigure}%
	\hfill
	\begin{subfigure}[b]{0.162\textwidth}
		\centering
		\includegraphics[width=\textwidth]{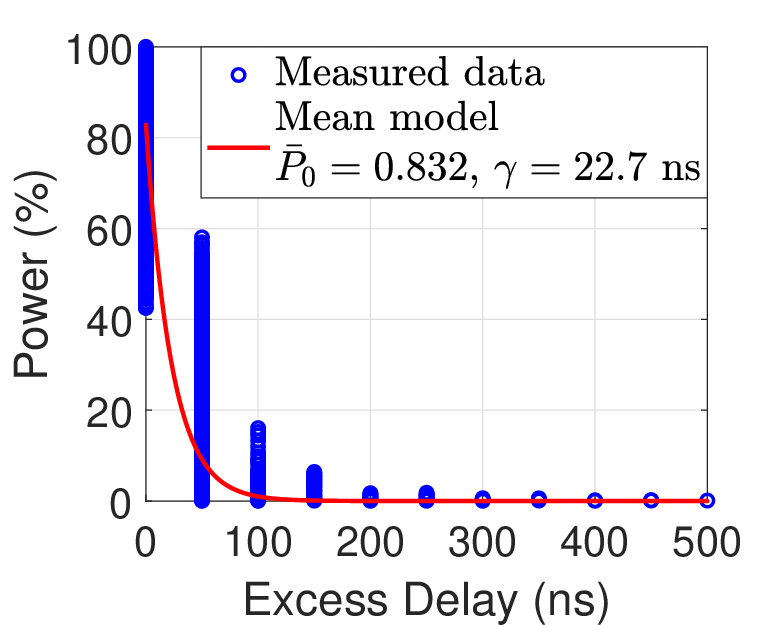}
		\caption{Day 2 Low Rx}
		\label{fig:day2low}
	\end{subfigure}
	\caption{Measured results and mean model for the multipath power versus excess delay.}
	\label{fig:multipath_power}
\end{figure}

Fig. \ref{fig:multipath_power} displays a comparison of three PDP datasets under different Rx heights and sea state conditions. It is observed from Fig. \ref{fig:multipath_power} that the mean power and decay time constant values alter little across the two days as well as antenna heights, with $\gamma$ in the range of around 23 ns to 25 ns. In contrast, the power decay constant is typically beyond 40 ns in terrestrial scenarios\cite{3GPP}. This suggests that received power often falls off more rapidly with excess delay in maritime channels, further corroborating the sparsity feature. 

\section{Conclusions}\label{sec_con}
\textcolor{black}{This paper has analyzed land-to-ship maritime wireless channels at 5.8 GHz, uncovering fundamental distinctions from terrestrial channels, such as multi-timescale fading (including SWIFT fading), channel sparsity, and pronounced environmental sensitivity. A series of models have been developed to comprehensively characterize the channel.} First, we have proposed a new large-scale path loss model, i.e., the dual-slope \textcolor{black}{CI-MTR} model, that enjoys solid physical foundation, low complexity, high accuracy, and good adaptability over a vast range of distances and different sea states. It is observed that windy sea environments arouse higher deviation in large-scale path loss. Second, the concept of SWIFT fading has been presented, and the proposed SWIFT fading model can align with measured data, particularly for small movements of the antenna. Third, unlike most terrestrial scenarios where Rician fading can well depict the small-scale fading, the asymmetric Laplace distribution is effective under calm sea conditions, while the TWDP model better captures the bimodality of fading behavior under adverse sea conditions. Furthermore, quantitative analysis based on the Gini index implies that the maritime propagation scenario possesses a higher degree of sparsity compared with typical terrestrial counterparts. \textcolor{black}{These findings advance the understanding of maritime propagation, offering valuable insights for system design in dynamic oceanic environments.}

\textcolor{black}{While the aforementioned models are validated for 5.8 GHz, their generalizability to other frequency bands or maritime environments remains an open question. Additionally, this study is limited to the specific sea states, ship type, and carrier bandwidth. Therefore, the} following aspects may be considered for future work:
	\begin{itemize}
		\item Conducting channel measurements and modeling across diverse sea areas \textcolor{black}{and frequency bands} to generalize current findings.
		\item Collecting more channel data and vessel motion status concurrently to enhance the SWIFT fading model.
		\item Performing extended-bandwidth channel measurements to experimentally validate the channel sparsity analysis.
	\end{itemize}



\section*{Appendix\\Proof of Lemma 1}
For the minimum resolvable multipath delay interval $\Delta\tau_2$, denote with $M$ the number of resolvable MPCs within a time delay bin of width $\Delta\tau_1$, then $M\geq1$. At a certain time instant $t$, the received signal $r_n(\tau)$ within the $n$-th delay bin ($n=1,...,N$) of width $\Delta\tau_1$ can be expressed as~\cite{Rappaport-WirelessCommunicationsPrinciplesandPractice}
\begin{equation}
	r_n(\tau) = \sum_{m=1}^{M}a_{n,m}\exp(j\varphi_{n,m})\delta(
	\tau-\tau_{n,m}),
	\label{eq_r_n}
\end{equation}

\noindent where $a_{n,m}$, $\varphi_{n,m}$, and $\tau_{n,m}$ are the amplitude, phase, and propagation time delay of the $m$-th resolvable MPC within the $n$-th delay bin, respectively. The notation $\delta(\cdot)$ stands for the Kronecker delta function. The received power within the $n$-th delay bin is
\begin{equation}
	\begin{split}
	P_n =&~\sum_{\tau=\tau_{n,1}}^{\tau_{n,M}}|r_n(\tau)|^2\\	=&~\sum_{\tau=\tau_{n,1}}^{\tau_{n,M}}\sum_{i=1}^{M}\sum_{k=1}^{M}a_{n,i}a_{n,k}\exp[j(\varphi_{n,i}-\varphi_{n,k})]\delta(\tau-\tau_{n,i})\\
	&~\cdot\delta(\tau-\tau_{n,k})\\
	=&~\sum_{i=1}^{M}\sum_{k=1}^{M}a_{n,i}a_{n,k}\exp[j(\varphi_{n,i}-\varphi_{n,k})]\delta(\tau_{n,i}-\tau_{n,k}).
	\label{eq_P_n}
	\end{split}
\end{equation}

\noindent Since the time delay differs for each resolvable MPC, the double sum in~\eqref{eq_P_n} is zero for $k\neq i$. The received power can thus be simplified to
\begin{equation}
	P_n=\sum_{i=1}^{M}a_{n,i}^2\exp[j(\varphi_{n,i}-\varphi_{n,i})]=\sum_{i=1}^{M}a_{n,i}^2.
	\label{eq_P_n2}
\end{equation}

\noindent Eq.~\eqref{eq_P_n2} shows that the received power within the $n$-th delay bin is equal to the sum of the powers of all the resolvable MPCs within the delay bin. 
Therefore, if assuming the power of the $n$-th delay bin is equally distributed among the $M$ MPCs within it, then the power of each of the $M$ MPCs is $\frac{P_n}{M}$. The Gini index considering all the $MN$ MPCs in all the $N$ delay bins is hence calculated as
\begin{equation}
	G_2 = 1-2\sum_{n=1}^{N}\sum_{m=1}^{M}\frac{P_n}{MP_\text{tot}}\cdot\frac{MN-m-M(n-1)+1/2}{MN},
	\label{G1}
\end{equation}

\noindent where $P_\text{tot}$ represents the total received power of all the $MN$ resolvable MPCs. On the other hand, the Gini index $G_1$ corresponding to the bandwidth $B_1$ is given by
\begin{equation}
	G_1 = 1-2\sum_{n=1}^{N}\frac{P_{n}}{P_{\text{tot}}}\cdot\frac{N-n+1/2}{N}.
	\label{eq_G1}
\end{equation}

\noindent Consequently, the difference between $G_2$ and $G_1$ is as follows
\begin{equation}
	\begin{split}
		G_2-G_1=&~2\sum_{n=1}^{N}\frac{P_{n}}{P_{\text{tot}}}\cdot\frac{N-n+\frac{1}{2}}{N}-\\&~2\sum_{n=1}^{N}\sum_{m=1}^{M}\frac{P_{n}}{MP_{\text{tot}}}\cdot\frac{MN-m-M(n-1)+\frac{1}{2}}{MN}\\
		=&~\frac{2}{NP_\text{tot}}\sum_{n=1}^{N}P_{n}\bigg[N-n+\frac{1}{2}-N+\frac{M+1}{2M}\\
		&~+(n-1)-\frac{1}{2M}\bigg]\\
		=&~0.
	\end{split}
\end{equation}

\noindent Thus, we obtain $G_2=G_1$.

\end{document}